\documentclass[aip,pof,preprint]{revtex4-1}

\usepackage[utf8]{inputenc}
\usepackage[T1]{fontenc}
\usepackage{graphicx}
\usepackage{amsmath}
\usepackage{amssymb}
\usepackage{array}
\usepackage{bm}
\usepackage{dcolumn}
\usepackage{comment}
\usepackage{etoolbox}
\usepackage{multirow}
\usepackage[normalem]{ulem}
\usepackage{xcolor}
\usepackage{booktabs}
\usepackage[ruled,linesnumbered]{algorithm2e}
\usepackage{ulem}
\DontPrintSemicolon
\AtBeginDocument{\IncMargin{4pt}}
\usepackage[hidelinks]{hyperref}
\usepackage{xurl}

\makeatletter
\def\@email#1#2{%
  \endgroup
  \patchcmd{\titleblock@produce}
    {\frontmatter@RRAPformat}
    {\frontmatter@RRAPformat{\produce@RRAP{*#1\href{mailto:#2}{#2}}}\frontmatter@RRAPformat}
    {}{}
}%
\makeatother

\begin{document}

\title{Extracting Single-Mode Rayleigh--Taylor Instability dynamics from data via SINDy}
% Data-Driven Identification of Single-Mode Rayleigh--Taylor Instability Dynamics Using SINDy

\author{Chuanxing Wang}
\affiliation{Graduate School, China Academy of Engineering Physics, Beijing 100088, China}

\author{Hui Luo}
\email{hluo@gscaep.ac.cn}
\affiliation{Graduate School, China Academy of Engineering Physics, Beijing 100088, China}

\author{Guohuai Zhu}
\affiliation{Beijing Computational Science Research Center, Beijing 100193, China}

\author{Hongbo Cai}
\affiliation{Graduate School, China Academy of Engineering Physics, Beijing 100088, China}
\affiliation{Beijing Computational Science Research Center, Beijing 100193, China}

\author{Shaoping Zhu}
\affiliation{Graduate School, China Academy of Engineering Physics, Beijing 100088, China}
\affiliation{Institute of Applied Physics and Computational Mathematics, Beijing 100094, China}

\author{Mingxing Luo}
\affiliation{Beijing Computational Science Research Center, Beijing 100193, China}

\date{\today}

% Local single spacing for the algorithm in the preprint layout.
\begingroup
\renewcommand{\baselinestretch}{1}\selectfont

\begin{abstract}
Rayleigh-Taylor instability (RTI) is of fundamental importance in fields such as inertial confinement fusion and supernova explosions, and its evolution is commonly described by potential--flow theory and buoyancy--drag models. However, both types of models struggle to accurately capture the overshoot behavior of bubble and spike velocities, as well as the subsequent reacceleration stage. To address these limitations, we develop a physics-constrained, data-driven framework based on the integral formulation of Sparse Identification of Nonlinear Dynamics (SINDy), which extracts low-dimensional dynamical equations for bubble and spike evolution directly from data. 
By introducing the nondimensional initial amplitude $A_0 = k\eta_0$ as an explicit parameter, we identify distinct dynamical equations for the bubble and the spike based on data spanning multiple Atwood numbers and initial amplitudes. The resulting low-dimensional dynamical equations not only reproduce the overshoot behavior of the velocity with good accuracy, but also capture the reacceleration stage well, while maintaining high predictive accuracy at unseen values of the wavenumber, acceleration, and Atwood number. This work provides a physics-constrained, data-driven framework complementary to conventional potential-flow and buoyancy--drag models, accurately describing the dynamical evolution of the bubble and spike at high Atwood numbers, and provides a new perspective for directly extracting the dynamical equations of interfacial instabilities from data. 
\end{abstract}
\maketitle

\maketitle %\maketitle must follow title, authors, abstract and \pacs

%==================================================================
\section{Introduction}
%==================================================================
Rayleigh-Taylor instability (RTI) refers to the unstable growth of small perturbations at the interface between two fluids of different densities when the acceleration points from the lighter fluid to the heavier fluid \cite{Rayleigh,Taylor1950instability}.
As the instability develops, the lighter fluid penetrates into the heavier fluid to form bubbles, whereas the heavier fluid penetrates into the lighter fluid to form spikes. 
RTI occurs in a broad range of scenarios, including inertial confinement fusion (ICF) \cite{2014NIFReview,2016Inertial} and supernova explosions \cite{cabot2006supernovae,zingale2005supernovae}.
During ICF implosions, the growth of RTI drives the mixing between the capsule ablator and fuel, degrading the spherical symmetry and compression efficiency of the implosion and thus severely compromising the ignition performance \cite{Clark2016NIF,Zhou2019Mixing,Srinivasan2014Mix}. 
A thorough theoretical investigation of RTI evolution is therefore of considerable scientific significance and engineering value, as it elucidates the underlying physical mechanisms and provides a theoretical basis for suppressing instability growth.

Existing theoretical descriptions of RTI are based primarily on potential-flow formulations {\color{black}\cite{Layzer1955,Goncharov2002,Zufiria1988,Kull1986,Mikaelian2003}} and buoyancy--drag models \cite{Dimonte2000,Oron2001}. 
Both approaches formulate evolution equations for the velocities of the bubble and spike tips, with the corresponding amplitude evolution obtained through temporal integration. 
Assuming incompressibility, inviscidity and irrotational flow, Layzer approximated the interfacial profile in the vicinity of the bubble tip by a paraboloid and derived an ordinary differential equation for the bubble velocity using an expansion of the velocity potential \cite{Layzer1955}. 
However, the model is restricted to the limiting case in which one fluid is replaced by vacuum, corresponding to $\mathit{A}_T=1$. 
Goncharov generalized the Layzer-type theory to arbitrary Atwood numbers, obtaining a continuous solution spanning the linear exponential-growth and nonlinear quasi-steady stages, and deriving the terminal velocities of two- and three-dimensional bubbles \cite{Goncharov2002}, thus extending the description into the nonlinear regime. 
Liu \textit{et al.} introduced a double-source velocity potential within the classical potential-flow framework, constructing a coupled dynamical system in terms of tip velocity, curvature, and source distance. 
This system successfully captures the long-term co-evolution of velocity and curvature \cite{Liu2023} and reproduces the velocity overshoot, in which the bubble velocity transiently exceeds and subsequently relaxes back toward its late-time asymptotic value. 
Nevertheless, its predictive accuracy for spike evolution at high Atwood numbers remains limited.
Building on this framework, Liu \textit{et al.} further introduced a bilaterally rotational flow system around the bubble and incorporated vorticity accumulation into the evolution equations, thereby capturing the evolution from the initial saturation of the quasi-steady stage to the reacceleration stage \cite{Liu2025}.
Within this framework, the reacceleration is attributed to the progressive breaking of the early buoyancy--drag equilibrium due to vorticity accumulation. 
Nevertheless, the model remains limited in its ability to predict spike dynamics.
{\color{black}More recently, Piriz \textit{et al.} \cite{Piriz2026} proposed a momentum-based description of the reacceleration stage that treats the bubble and the spike on the same footing: the buoyancy force is taken as the sole source of momentum, and vorticity merely drains momentum from the translational motion into the rotational motion, thereby arresting the lateral expansion (contraction) of the bubble (spike). Once the lateral expansion is arrested at a threshold amplitude \cite{Piriz2024}, the participating mass becomes linear in the amplitude and the momentum equation admits a constant-acceleration solution, which reproduces the reacceleration of the interface
without introducing any additional force.}

Buoyancy--drag models, in contrast, interpret interfacial motion as a competition between buoyancy driven by the density difference and inertial drag associated with momentum exchange, prescribing the specific forms of the buoyancy and drag terms directly via assumption. 
Based on a uniform approximation within the bubble and spike regions and mass conservation, Dimonte established a buoyancy--drag model for the mixing zone \cite{Dimonte2000}; Oron \textit{et al.} further incorporated coefficients derived from potential-flow theory into this buoyancy--drag model to describe the late-time scaling of both Rayleigh--Taylor and Richtmyer--Meshkov instabilities across different dimensionalities \cite{Oron2001}.

Potential-flow theory relies on a specific irrotational assumption and yields poor predictive accuracy for spikes at high Atwood numbers.
Classical buoyancy--drag models, in turn, treat the coefficients of buoyancy and drag terms as independent of the instantaneous amplitude $A$ and initial amplitude $A_0$, thereby failing to accurately capture the effects of $A_0$ on nonlinear evolution. Furthermore, given that both the magnitude and the onset time of the velocity overshoots strongly depend on the initial amplitude, these models cannot reproduce the overshoot phenomenon. 
In addition, classical buoyancy--drag models fail to recover the linear limit.

To address these limitations, we propose a data-driven dynamical modeling framework for interfacial instabilities that employs the integral-form Sparse Identification of Nonlinear Dynamics (SINDy) method \cite{Brunton2016,Rudy2017PDEFind,Champion2019Coordinates,Brunton2020MLFluid} to infer low-dimensional dynamical equations from RTI data. 
The central idea of SINDy is as follows: given data on the state variables and their time derivatives, one first constructs a candidate library containing polynomial, trigonometric, rational, or physics-informed basis functions, and then uses sparse regression to select a small subset of dominant terms that constitute an explicit governing equation.
Compared with other machine-learning approaches, SINDy-derived models are generally more interpretable and can directly reveal the relative contributions of different physical mechanisms to dynamical evolution. 
In contrast to traditional theoretical derivations, SINDy can automatically extract effective dynamical forms from the data even when the prior assumptions are incomplete.
In the present work, we infer separate low-dimensional dynamical equations governing bubble and spike evolution from direct numerical simulation (DNS) data. 
Building on the classical buoyancy--drag framework, we employ the integral formulation of SINDy to identify state- and parameter-dependent buoyancy and drag coefficients, with explicit dependence on the Atwood number, the instantaneous amplitude, and the initial perturbation amplitude. 
The resulting equations accurately reproduce the overshoot--relaxation behavior of the bubble and spike velocities and capture their full temporal evolution, including the late-time reacceleration stage. 
To the best of our knowledge, this is the first application of an integral-form SINDy framework with physical constraints from the buoyancy–drag structure to extract reduced-order dynamical equations for single-mode Rayleigh–Taylor instability. 
The buoyancy–drag structure is imposed as a physically motivated prior, while the state- and parameter-dependent closure functions are identified from DNS data
Because the fluids are treated as incompressible, an incompressible formulation is adopted throughout the simulations.

The remainder of this paper is organized as follows. Section~\ref{sec:Methods} presents the formulation of the problem and its nondimensionalization, introduces the physics-constrained, data-driven modeling framework for RTI, and describes the datasets used for model training and evaluation. Section~\ref{sec:results} presents the low-dimensional dynamical equations identified for the bubble and spike using the integral formulation of SINDy, together with the corresponding generalization performance and related analyses. Section~\ref{sec:conclusion} summarizes the main findings and discusses directions for future work.

%==================================================================
\section{Methods}
%==================================================================
\label{sec:Methods}
\subsection{Problem formulation and nondimensionalization}

We consider a two-dimensional, incompressible, single-mode Rayleigh--Taylor instability (RTI) subject to constant acceleration $g$. 
The initial interfacial disturbance is characterized by wavenumber $k$ and amplitude $\eta_0$. 
The densities of the heavy and light fluids are denoted by $\rho_h$ and $\rho_l$, respectively, and the corresponding Atwood number is defined as
\begin{equation}
    \mathit{A}_T
    = \frac{\rho_h-\rho_l}{\rho_h+\rho_l}.
\end{equation}

\begin{figure}[htbp]
\centering
\includegraphics[width=3.4039in]{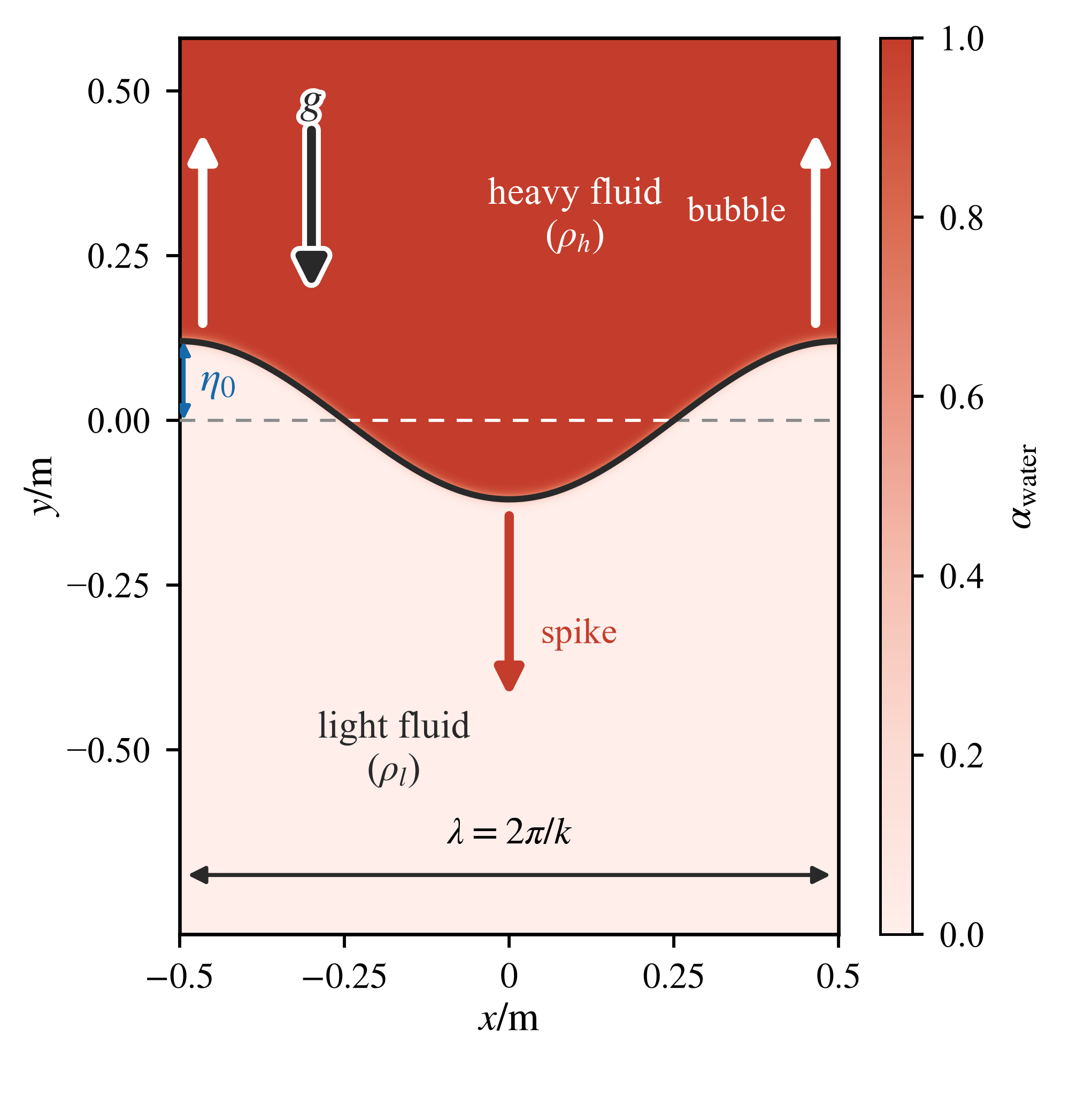}
\caption{Configuration of the single-mode Rayleigh--Taylor instability considered in
this work. The heavy fluid ($\rho_h$) rests above the light fluid ($\rho_l$)
with $\bm{g}=(0,-g,0)$, so the light fluid rises as the bubble and the
heavy fluid descends as the spike. The panel spans one wavelength in
$x$ (lateral boundaries periodic). Following
the deposited initialisation, $y_{\mathrm{int}}(x)=-\eta_0\cos(2\pi x/\lambda)$:
the spike tip lies at $x=0$ and the bubble tips at $x=\pm\lambda/2$.}
\label{fig:config}
\end{figure}

The problem geometry is sketched in Fig.~\ref{fig:config}. A heavy fluid
($\rho_h$) rests above a light fluid ($\rho_l$) with $\bm{g}=(0,-g,0)$, so the
interface is unstable: the light fluid rises as the \emph{bubble} and the
heavy fluid descends as the \emph{spike}. The interface is perturbed by a
single mode, $\eta(x,0)=-\eta_0\cos(2\pi x/\lambda)$ with $\lambda=2\pi/k$; the
lateral boundaries are periodic and span exactly one wavelength. The two
fluids are distinguished by the phase fraction $\alpha_{\mathrm{water}}$.

The following dimensionless variables are introduced:
\begin{equation}
    \tau = t\sqrt{gk}, \quad
    A = k\eta, \quad
    U = \frac{V}{\sqrt{g/k}}, \quad
    A_0 = k\eta_0.
    \label{eq:dimensionless_variables}
\end{equation}
where $\eta(t)$ and $V(t)=\mathrm{d}\eta/\mathrm{d}t$ denote the dimensional amplitude and tip velocity (bubble/spike), respectively, and $A_0$ is the dimensionless initial amplitude. 
In what follows, all data preprocessing, SINDy identification, and forward integration are performed using the nondimensional variables in Eq.~\eqref{eq:dimensionless_variables} (in particular, $\tau$ and $U$). 
For visualization only, we additionally introduce the rescaled plotting coordinate 
\begin{equation}
\theta=t\sqrt{\mathit{A}_T\,gk}=\tau\,\sqrt{\mathit{A}_T},
\label{eq:theta}
\end{equation}
which enables comparison across different Atwood numbers at approximately the corresponding instability stages. 
The coordinate $\theta$ is used only for plotting and does not enter the regression or identified dynamical equations.

Under this nondimensionalization, $g$ and $k$ no longer appear explicitly in the governing equations, and the dynamics are fully described by $(\tau, \mathit{A}_T, A, U, A_0)$. 
Taking $A$ and $U$ as state variables, the motion of the bubble and spike tips is governed by the following two-dimensional dynamical equation:
\begin{equation}
    \frac{\mathrm{d}U}{\mathrm{d}\tau} = f(U,\,A,\,A_0,\,\mathit{A}_T), \qquad
    \frac{\mathrm{d}A}{\mathrm{d}\tau} = U,
    \label{eq:system}
\end{equation}
The second equation follows identically from the definition of $U$, while the main objective of the present study is to identify the function $f$ in the first equation from the available data.

\subsection{Data-driven modeling framework}

We propose a physics-constrained, data-driven framework for Rayleigh--Taylor instability (RTI) that infers low-dimensional dynamical equations governing bubble and spike evolution directly from the corresponding amplitude and velocity time series. 
We construct a candidate-function library comprising functions of the state variables $A$ and $U$, the initial amplitude $A_0$, and the Atwood number $\mathit{A}_T$. Within the SINDy framework, sparse regression is then used to identify a parsimonious set of active terms in the governing function $f$. 
In the present RTI problem, an unconstrained search over a broad candidate-function library may produce insufficiently sparse models that are difficult to interpret physically. 
To reduce the search space and improve the physical interpretability, we parameterize the unknown acceleration term $f$ in Eq.~\eqref{eq:system} using the following buoyancy--drag form:
\begin{equation}
    \begin{aligned}
        \frac{\mathrm{d}A}{\mathrm{d}\tau} &= U, \\
        \frac{\mathrm{d}U}{\mathrm{d}\tau}
        &= \mathit{A}_T\,A\left[1+B(A;A_0,\mathit{A}_T)\right]
        - D(A;A_0,\mathit{A}_T)U^2,
    \end{aligned}
    \label{eq:bd_sindy}
\end{equation}
where $B(A;A_0,\mathit{A}_T)$ represents the nonlinear correction relative to the baseline buoyancy term $\mathit{A}_T\,A$, and $D(A;A_0,\mathit{A}_T)$ is the effective drag coefficient. 
The functions $B$ and $D$ can be expressed as linear combinations of candidate basis functions:
\begin{equation}
    \begin{aligned}
        B(A;A_0,\mathit{A}_T)
        &= \sum_{j=1}^{N_B}
        \beta_j\,\phi_j^{(B)}(A;A_0,\mathit{A}_T), \\
        D(A;A_0,\mathit{A}_T)
        &= \sum_{\ell=1}^{N_D}
        \gamma_\ell\,\phi_\ell^{(D)}(A;A_0,\mathit{A}_T).
    \end{aligned}
    \label{eq:bd_expansion}
\end{equation}
Here, $\phi_j^{(B)}$ and $\phi_\ell^{(D)}$ denote the candidate basis functions for the buoyancy correction and drag coefficient, respectively, and $\beta_j$ and $\gamma_\ell$ are the coefficients to be identified. 
The corresponding libraries contain $N_B$ and $N_D$ terms. 
{\color{black}The buoyancy library is further restricted by the linear limit. Since Eq.~\eqref{eq:bd_sindy} should reduce to the classical linear RTI equation as $A\to0$ and $A_0\to0$, the pure $\mathit{A}_T$-monomials are excluded from the buoyancy library so that $B(0;0,\mathit{A}_T)=0$ holds identically; the drag library is left unrestricted, as $DU^2$ is $O(U^2)$ and does not enter the linear limit. The formal definition is given in Appendix~\ref{app:function}.}
The training set comprises $N_{\mathrm{train}}$ independent RTI cases. Each case $r$ has a prescribed initial amplitude $A_0^{(r)}$ and Atwood number $\mathit{A}_T^{(r)}$. 
The corresponding state variables $A^{(r)}(\tau)$ and $U^{(r)}(\tau)$ are sampled at $N_r$ discrete times $\{\tau_i^{(r)}\}_{i=0}^{N_r-1}$. 
The values of the candidate basis functions along this trajectory are denoted by
\begin{equation}
    \begin{aligned}
        \phi_j^{(B,r)}(\tau)
        &\equiv \phi_j^{(B)}
        \left(A^{(r)}(\tau);A_0^{(r)},\mathit{A}_T^{(r)}\right), \\
        \phi_\ell^{(D,r)}(\tau)
        &\equiv \phi_\ell^{(D)}
        \left(A^{(r)}(\tau);A_0^{(r)},\mathit{A}_T^{(r)}\right).
    \end{aligned}
    \label{eq:trajectory_library}
\end{equation}
Here, the superscript $r$ identifies the corresponding training case.

Compared with the classical buoyancy--drag model, the present formulation allows the effective buoyancy and drag contributions to depend on the instantaneous amplitude, the initial amplitude, and the Atwood number. It is intended to describe the full evolution of the interface, covering the linear stage, the nonlinear saturation of the bubble and spike velocities, and the subsequent reacceleration stage.
In the small-amplitude regime ($A,U\ll1$), the effective buoyancy correction $\mathit{A}_T\,A B(A;A_0,\mathit{A}_T)$ and the drag contribution $D(A;A_0,\mathit{A}_T)U^2$ remain subdominant to the baseline buoyancy term $\mathit{A}_T\,A$. Equation~\eqref{eq:bd_sindy} therefore reduces, to leading order, to
\begin{equation}
    \frac{\mathrm{d}U}{\mathrm{d}\tau}
    \approx
    \mathit{A}_T\,A.
\end{equation}
Upon restoring dimensions, this relation becomes
\begin{equation}
    \frac{\mathrm{d}^2\eta}{\mathrm{d}t^2}
    =
    \frac{\mathrm{d}V}{\mathrm{d}t}
    \approx
    \mathit{A}_T\,gk\eta,
\end{equation}
which is consistent with the classical linear RTI equation.

In the differential formulation of SINDy, the acceleration $\mathrm{d}U/\mathrm{d}\tau$ must be estimated by numerically differentiating the velocity time series. 
Because numerical differentiation amplifies high-frequency noise and discretization errors, this step may compromise the numerical robustness of the sparse-regression procedure. 
To mitigate this sensitivity, we adopt an  integral formulation of SINDy \cite{Schaeffer2017,Messenger2021}. 
Integrating Eq.~\eqref{eq:bd_sindy} over the interval between two adjacent sampling times, $\tau_i^{(r)}$ and $\tau_{i+1}^{(r)}$, we define the regression target for a single time interval as
\begin{equation}
    y_{r,i}
    \equiv
    U_{i+1}^{(r)} - U_i^{(r)}
    -
    \mathit{A}_T^{(r)}
    \int_{\tau_i^{(r)}}^{\tau_{i+1}^{(r)}}
    A^{(r)}(\tau)\,\mathrm{d}\tau .
    \label{eq:interval_target}
\end{equation}
The quantity $y_{r,i}$ denotes the regression target associated with time interval $i$ of case $r$. Using the expansions in Eq.~\eqref{eq:bd_expansion}, we obtain
\begin{align}
    y_{r,i}
    &=
    \sum_{j=1}^{N_B}
    \mathit{A}_T^{(r)}\beta_j
    \int_{\tau_i^{(r)}}^{\tau_{i+1}^{(r)}}
    A^{(r)}(\tau)
    \phi_j^{(B,r)}(\tau)\,\mathrm{d}\tau
    \notag \\
    &\quad-
    \sum_{\ell=1}^{N_D}
    \gamma_\ell
    \int_{\tau_i^{(r)}}^{\tau_{i+1}^{(r)}}
    \left[U^{(r)}(\tau)\right]^2
    \phi_\ell^{(D,r)}(\tau)\,\mathrm{d}\tau
    +
    \varepsilon_{r,i},
\label{eq:integral_sindy}
\end{align}
where $\varepsilon_{r,i}$ accounts for the model discrepancy, including errors introduced by numerical quadrature and other sources.

The entries of the interval-integrated feature matrices associated with the buoyancy and drag contributions are defined as
\begin{equation}
    \Theta_{(r,i),j}^{(B)}
    \equiv
    \mathit{A}_T^{(r)}\int_{\tau_i^{(r)}}^{\tau_{i+1}^{(r)}}
    A^{(r)}(\tau)
    \phi_j^{(B,r)}(\tau)\,\mathrm{d}\tau ,
\label{eq:buoyancy_features}
\end{equation}
\begin{equation}
    \Theta_{(r,i),\ell}^{(D)}
    \equiv
    -
    \int_{\tau_i^{(r)}}^{\tau_{i+1}^{(r)}}
    \left[U^{(r)}(\tau)\right]^2
    \phi_\ell^{(D,r)}(\tau)\,\mathrm{d}\tau .
\label{eq:drag_features}
\end{equation}

Across the $N_{\mathrm{traj}}$ training trajectories, the total number of sampling intervals is $N=\sum_{r=1}^{N_{\mathrm{traj}}}(N_r-1)$, and the candidate library contains $p=N_B+N_D$ terms. Stacking the scalar targets $y_{r,i}$ in the same order as the rows of the feature matrix gives the target vector $\bm{y}=\operatorname{stack}\{y_{r,i}\}\in\mathbb{R}^{N}$. The buoyancy and drag coefficients are collected into a single vector:
\begin{equation}
    \bm{c}
    =
    \begin{bmatrix}
        \beta_1,\ldots,\beta_{N_B},
        \gamma_1,\ldots,\gamma_{N_D}
    \end{bmatrix}^{\mathsf{T}}
    \in\mathbb{R}^{p}.
    \label{eq:coefficient_vector}
\end{equation}

The design matrix for the integral SINDy formulation is formed by horizontally concatenating the buoyancy and drag blocks:
\begin{equation}
    \bm{\Theta}
    =
    \left[
        \bm{\Theta}^{(B)}
        \;\;
        \bm{\Theta}^{(D)}
    \right]
    \in\mathbb{R}^{N\times p},
    \qquad
    \label{eq:design_matrix}
\end{equation}
where
\[
    \bm{\Theta}^{(B)}\in\mathbb{R}^{N\times N_B},
    \qquad
    \bm{\Theta}^{(D)}\in\mathbb{R}^{N\times N_D}.
\] 
In each block, rows correspond to sampling intervals and columns correspond to candidate basis functions. 
Accordingly, Eq.~\eqref{eq:integral_sindy} can be written as a linear regression problem for the unknown coefficients:
\begin{equation}
    \bm{y}
    =
    \bm{\Theta}\bm{c}
    +
    \bm{\varepsilon},
    \label{eq:weak_regression}
\end{equation}
where $\bm{\varepsilon}\in\mathbb{R}^{N}$ is the stacked residual vector. 
To reduce the sensitivity of the regression to differences in column magnitude, we normalize each column of the integral-form design matrix $\bm{\Theta}$ to unit $L_2$ norm. For $m=1,\ldots,p$, let
\begin{equation}
    s_m
    =
    \left\|
        \bm{\Theta}_{:,m}
    \right\|_2,
    \qquad
    m=1,\ldots,p,
    \label{eq:column_norm}
\end{equation}
and define
\begin{equation}
    \bm{S}
    =
    \operatorname{diag}(s_1,\ldots,s_p),
    \qquad
    \widetilde{\bm{\Theta}}
    =
    \bm{\Theta}\bm{S}^{-1}.
    \label{eq:column_scaling}
\end{equation}

To preserve the regression relation, we define the rescaled coefficient vector as
\begin{equation}
    \widetilde{\bm{c}}
    =
    \bm{S}\bm{c},
    \qquad
    \bm{\Theta}\bm{c}
    =
    \widetilde{\bm{\Theta}}\widetilde{\bm{c}}.
    \label{eq:normalized_coefficients}
\end{equation}
Using the normalized design matrix, we first solve the following ridge-regularized least-squares problem:
\begin{equation}
    \widetilde{\bm{c}}^{(0)}
    =
    \underset{\bm{z}\in\mathbb{R}^{p}}{\arg\min}
    \left\{
        \frac{1}{2}
        \left\|
            \widetilde{\bm{\Theta}}\bm{z}-\bm{y}
        \right\|_2^2
        +
        \frac{\lambda_r}{2}
        \left\|\bm{z}\right\|_2^2
    \right\},
    \qquad
    \lambda_r=10^{-8}.
    \label{eq:ridge}
\end{equation}
We then obtain a sparse model using ridge-regularized sequentially thresholded least squares (Ridge-STLSQ). At each iteration, we evaluate the rescaled coefficient magnitudes $I_m=|\widetilde{c}_m|$ and remove candidate terms with $I_m<\delta_I$. The coefficients of the retained terms are refitted by ridge regression, and this procedure is repeated until the active set is unchanged between two successive iterations. The optimal threshold can be selected from a candidate set $\Delta_I$ by minimizing the training root-mean-square errors (RMSE) plus a penalty on the number of active terms:
\begin{equation}
    \delta_I^\ast
    =
    \underset{\delta_I\in\Delta_I}{\arg\min}
    \left[
        \operatorname{RMSE}_{\mathrm{train}}(\delta_I)
        +
        \lambda_s N_{\mathrm{active}}(\delta_I)
    \right],
    \label{eq:threshold_score}
\end{equation}
where $N_{\mathrm{active}}$ is the number of terms retained after convergence, and $\lambda_s$ controls the strength of the sparsity penalty. We recover the original coefficients as $\bm{c}^{\ast}=\bm{S}^{-1}\widetilde{\bm{c}}(\delta_I^\ast)$ and substitute this coefficient vector into Eq.~\eqref{eq:bd_sindy} to obtain the identified dynamical system. Further details of threshold selection are provided in the Appendix~\ref{app:threshold}, and the complete identification procedure is presented in Algorithm~\ref{alg:sindy}.

\begin{algorithm}[H]
%\begin{algorithm*}[t]
\small
\caption{Identification of RTI dynamics using integral-form SINDy}
\label{alg:sindy}

\KwIn{
Training datasets
$\mathcal{D}_{\mathrm{train}}$ ;
candidate library $\{\phi_j^{(B)}\}_{j=1}^{N_B}$
for buoyancy correction;
candidate library $\{\phi_\ell^{(D)}\}_{\ell=1}^{N_D}$
for drag;
candidate threshold set $\Delta_I$;
ridge regularization parameter $\lambda_r$;
sparsity penalty parameter $\lambda_s$
}

\KwOut{
Sparse coefficient vector $\bm{c}^{\ast}$ and
identified dynamical equations for bubble and spike evolution
}

For each training case, construct the interval targets and
integrated candidate features\;

Combine all training cases to obtain the integral regression system
$\bm{y}=\bm{\Theta}\bm{c}+\bm{\varepsilon}$\;

Compute $s_m=\lVert\bm{\Theta}_{:,m}\rVert_2$,
$m=1,\ldots,p$,
and define $\mathcal{J}=\{m:s_m>0\}$\;

Set $s_m=1$ for $m\notin\mathcal{J}$,
define $\bm{S}=\operatorname{diag}(s_1,\ldots,s_p)$,
and normalize
$\widetilde{\bm{\Theta}}=\bm{\Theta}\bm{S}^{-1}$\;

\For{$\delta_I\in\Delta_I$}{

    Initialize the active set
    $\mathcal{S}\gets\mathcal{J}$\;

    \Repeat{the active set $\mathcal{S}$ remains unchanged}{

        Set
        $\widetilde{\bm{c}}\gets\bm{0}\in\mathbb{R}^{p}$\;

        \If{$\mathcal{S}\neq\varnothing$}{

            Fit ridge regression with parameter $\lambda_r$
            on $\widetilde{\bm{\Theta}}_{:,\mathcal{S}}$
            to obtain
            $\widetilde{\bm{c}}_{\mathcal{S}}$\;

            Compute the normalized coefficient magnitudes
            $I_m=|\widetilde{c}_m|$,
            $m\in\mathcal{S}$,
            and update
            $\mathcal{S}\gets
            \{m\in\mathcal{S}:I_m\geq\delta_I\}$\;
        }
    }

    Recover and store
    $\bm{c}(\delta_I)=
    \bm{S}^{-1}\widetilde{\bm{c}}$\;

    Evaluate
    $\operatorname{RMSE}_{\mathrm{train}}(\delta_I)$
    using $\bm{c}(\delta_I)$ on
    $\mathcal{D}_{\mathrm{train}}$\;

    Set
    $N_{\mathrm{active}}(\delta_I)=|\mathcal{S}|$
    and compute
    $\operatorname{Score}(\delta_I)=
    \operatorname{RMSE}_{\mathrm{train}}(\delta_I)
    +\lambda_s N_{\mathrm{active}}(\delta_I)$\;
}

Select
$\delta_I^\ast=
\underset{\delta_I\in\Delta_I}{\arg\min}
\operatorname{Score}(\delta_I)$
and set
$\bm{c}^{\ast}=\bm{c}(\delta_I^\ast)$\;

Reconstruct $B$ and $D$ from $\bm{c}^{\ast}$ and
substitute them into the buoyancy--drag equations\;

\KwRet{
$\bm{c}^{\ast}$ and the resulting
bubble and spike dynamical equations
}\;

\end{algorithm}
%\end{algorithm*}

\subsection{Training and Test Datasets}
\label{sec:dataset_split}

For the training dataset, the gravitational acceleration and wavenumber were fixed at $g=9.8\,\mathrm{m/s^2}$ and $k=2\pi\,\mathrm{m^{-1}}$, respectively.
The Atwood number and initial perturbation amplitude were selected as
$$
    \mathit{A}_T
    \in
    \{0.1,\,0.3,\,0.5,\,0.7,\,0.9\}
$$
and
$$
    \eta_0
    \in
    \{0.01,\,0.03,\,0.05,\,0.08,\,0.10,\,0.12,\,0.15\}\,\mathrm{m},
$$
respectively. Direct numerical simulations of single-mode RTI were performed for all combinations of these parameters,
yielding a total of $35$ training cases.
For each simulation case, the temporal evolutions of the bubble- and spike-tip amplitudes, $\eta_b(t)$ and $\eta_s(t)$, as well as their
corresponding velocities, $V_b(t)$ and $V_s(t)$, were extracted from the
direct numerical simulation data. These quantities were nondimensionalized according to Eqs.~\eqref{eq:dimensionless_variables}, yielding the dimensionless variables
$(\tau,A,U,A_0)$. Using these dimensionless data, the regression system was
formulated in the interval-integral form defined by
Eq.~\eqref{eq:integral_sindy}. In contrast to a differential formulation,
this approach does not require direct numerical differentiation of the
discrete velocity data and is therefore less susceptible to the
amplification of numerical noise.

To assess the generalizability of the identified dynamical equations to
parameter regimes not represented in the training data, three independent
test datasets were constructed.

The first test dataset was designed to evaluate generalization across
Atwood numbers. With the gravitational acceleration and wavenumber fixed at
$g=9.8\,\mathrm{m/s^2}$ and $k=2\pi\,\mathrm{m^{-1}}$, respectively, four Atwood numbers that were not included in the training dataset were considered: 
$$
    \mathit{A}_T
    \in
    \{0.2,\,0.4,\,0.6,\,0.8\},
$$
in combination with the seven initial perturbation amplitudes used in the training set, yielding 28 cases in total. This dataset was used to assess the model's interpolative generalization to previously unseen Atwood numbers.

The second test set was constructed to assess the model's generalization across gravitational accelerations. With the wavenumber fixed at $k=2\pi\,\mathrm{m^{-1}}$ and the Atwood number fixed at $A_T=0.5$, we considered three gravitational accelerations absent from the training data,
\[
    g\in\{8,\,10,\,12\}\,\mathrm{m\,s^{-2}},
\]
in combination with seven initial perturbation amplitudes,
yielding 21 cases in total. This test set was used to evaluate the predictive
performance of the identified model under gravitational accelerations outside
the training range.

The third test dataset was designed to evaluate generalization across
wavenumbers. With the gravitational acceleration fixed at
$g=9.8\,\mathrm{m/s^2}$ and the Atwood number fixed at $A_T=0.5$, two wavenumbers not encountered during training,
$$
    k\in\{\pi,\,3\pi\}\,\mathrm{m^{-1}},
$$
were considered together with seven initial perturbation amplitudes,
{\color{black}yielding 14 cases in total}. {\color{black}The wavenumber $k=2\pi\,\mathrm{m^{-1}}$, at which
every training case was generated, is excluded from this extrapolation test and
appears in the presentation of the results only as a reference.} This dataset
was employed to assess the model's ability to generalize across different
perturbation length scales.

None of the three test datasets was used in identifying the coefficients of
the dynamical equations; instead, they were reserved exclusively for 
independent evaluation of the generalization capability
and predictive accuracy of the identified model.

%==================================================================
\section{Results and Discussion}
%==================================================================
\label{sec:results}

In figure~\ref{fig:classic_failure}, DNS data are compared with predictions from the classical buoyancy--drag model, Liu's modified potential-flow theory, and the SINDy-identified low-dimensional dynamical model.
%In the classical buoyancy--drag model $dU/d\tau=C_a-C_dU^2$, where the buoyancy coefficient $C_a$ and the drag coefficient $C_d$ do not explicitly depend on either the instantaneous amplitude $A$ or the initial amplitude $A_0$.
{\color{black}In the classical buoyancy--drag model \cite{Dimonte2000,Oron2001}, the interfacial motion is described by a balance among buoyancy and quadratic drag,}
{\color{black}
\begin{equation}
    \begin{aligned}
        \left[(1-\mathit{A}_T)+C_a\,(1+\mathit{A}_T)\right]\frac{\mathrm{d}U}{\mathrm{d}\tau}
        &= 2\mathit{A}_T - C_d\,(1+\mathit{A}_T)\,U^2 \quad \text{(bubble)},\\[4pt]
        \left[(1+\mathit{A}_T)+C_a\,(1-\mathit{A}_T)\right]\frac{\mathrm{d}U}{\mathrm{d}\tau}
        &= 2\mathit{A}_T - C_d\,(1-\mathit{A}_T)\,U^2 \quad \text{(spike)},
    \end{aligned}
    \label{eq:classic_bd}
\end{equation}
}
{\color{black}where the coefficient $C_a$ accounts for the added mass and $C_d$ is the drag coefficient; both are constants that do not depend on the instantaneous amplitude $A$ or on the initial amplitude $A_0$.}
A global fit of the model coefficients to the DNS data from the entire training dataset yielded $C_a=2.64$, $C_d=3.25$ for bubble and $C_a=3.15$, $C_d=2.43$ for spike. 
Across the compared datasets, the mean RMSE of the dimensionless tip
velocity $U$ between the mode prediction and the DNS data was $0.063$ for the
bubble and $0.155$ for the spike.
The classical buoyancy--drag model predicts saturation of the tip velocity. 
But it doesn't accurately reproduce the exponential growth in the linear regime or the subsequent overshoot and reacceleration. 
Liu's modified potential-flow theory incorporates the evolution of tip curvature and partially reproduces the velocity overshoot of both bubbles and spikes. 
However, the predicted velocity for the high-Atwood-number spike case shown in the figure deviates markedly from DNS, particularly during the reacceleration stage. 
The mean RMSE of the dimensionless tip velocity $U$ between the mode prediction and the DNS data was $0.063$ for the bubble and $0.181$ for the spike.

\begin{figure*}[htbp]
\centering
\includegraphics[width=\textwidth]{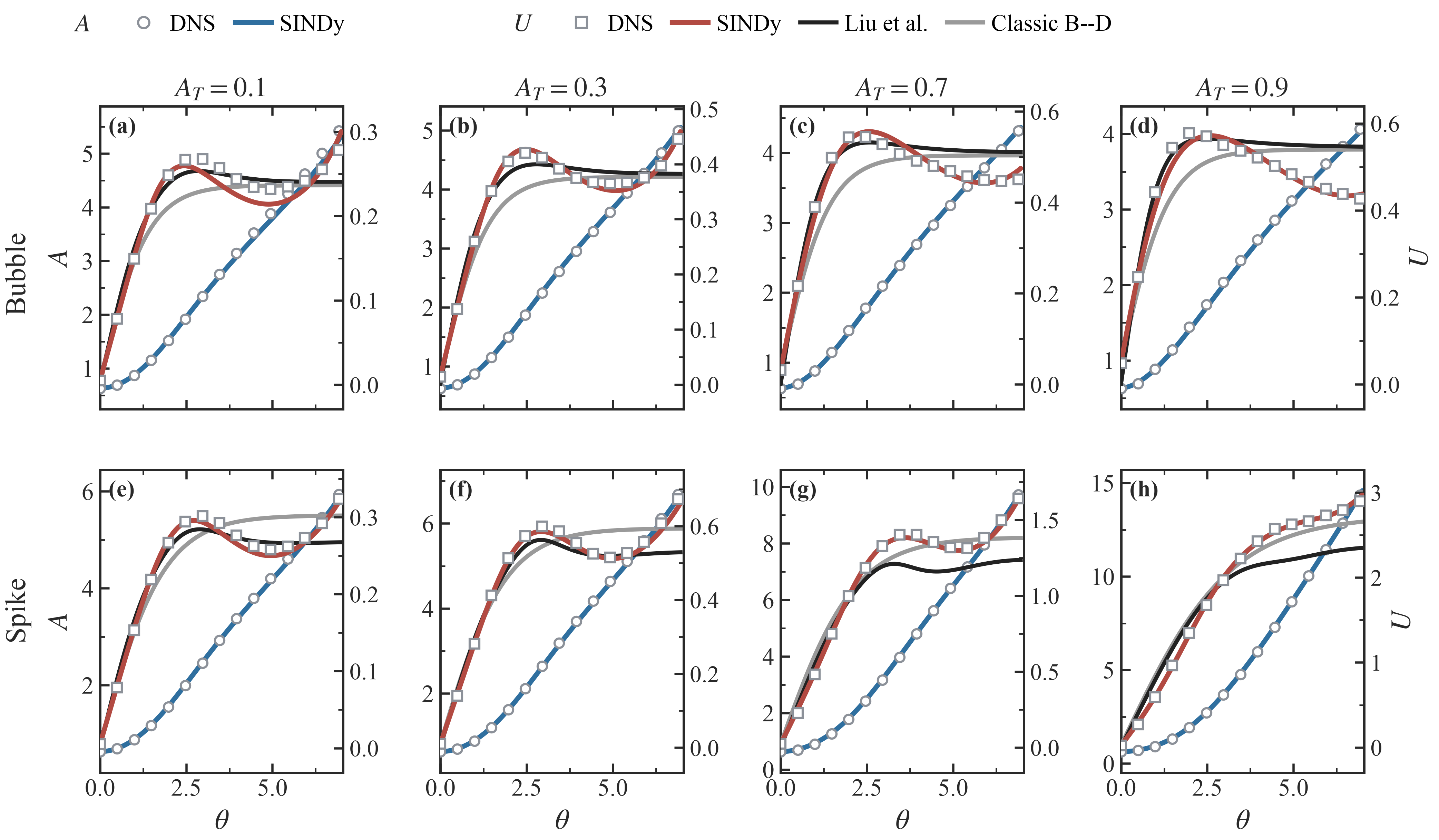}
\caption{Comparison of Liu's modified potential-flow theory, the classical buoyancy--drag model, and the SINDy-identified dynamical model with the DNS data. The first and second rows correspond to the bubble and spike, respectively, while the first through fourth columns correspond to $\mathit{A}_T=0.1$, $0.3$, $0.7$, and $0.9$, respectively. 
The initial amplitude is fixed at $\eta_0=0.1\,\mathrm{m}$ for all cases. 
}
    \label{fig:classic_failure}
\end{figure*}

Subject to the buoyancy--drag structure, the integral-form SINDy method was employed to identify the buoyancy-correction function $B(A;A_0,\mathit{A}_T)$ and the drag-coefficient function $D(A;A_0,\mathit{A}_T)$ from a prescribed candidate-function library. 
Both functions were allowed to vary with the instantaneous amplitude $A$, the initial amplitude $A_0$, and the Atwood number $\mathit{A}_T$. 
Figure~\ref{fig:a0_phase} presents the dimensionless phase-space trajectories obtained from DNS at a fixed Atwood number of $\mathit{A}_T=0.3$ for different values of $A_0$.
The trajectories for different initial amplitudes exhibit distinct peak velocities and peak times. Furthermore, these trajectories may intersect for different $A_0$.
Such intersections imply that the projected state $(A,U;\mathit{A}_T)$ does not uniquely determine the subsequent dynamics: in an identical $(A,U;\mathit{A}_T)$, trajectories originating from different $A_0$ may exhibit different accelerations $dU/d\tau$.
The dependence of the phase-space trajectory on the initial amplitude demonstrates that the dynamics cannot be represented by a single-valued autonomous closure formulated solely in terms of $(A,U;\mathit{A}_T)$. 
To retain the dependence of subsequent evolution on the initial condition, dynamical equations must explicitly include the initial amplitude $A_0$ or an equivalent state descriptor. 
This finding provides both physical and mathematical justification for incorporating $A_0$-dependent terms into the candidate-function library{\color{black}; the cost of omitting them is quantified by the ablation study in Appendix~\ref{app:a0_ablation}}.

\begin{figure*}[htbp]
\centering
\includegraphics[width=\textwidth]{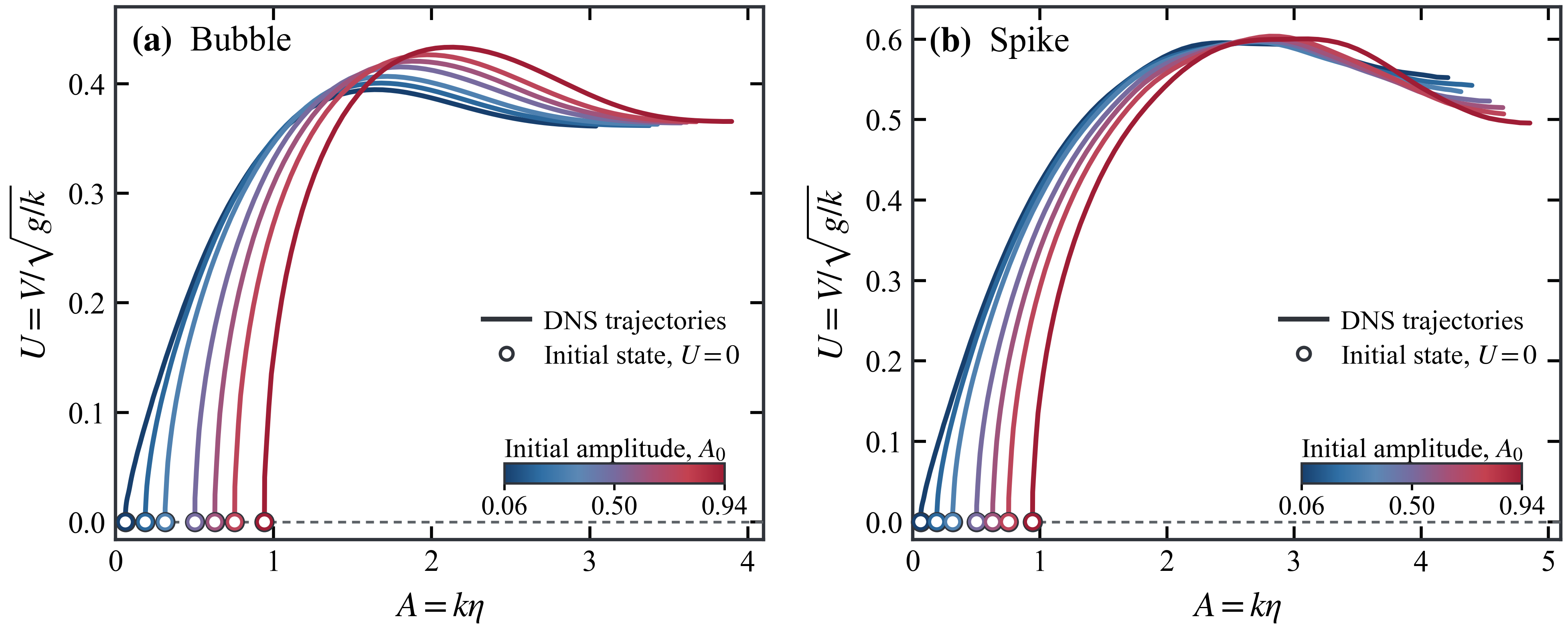}
\caption{DNS-derived dimensionless phase trajectories for different initial amplitudes at a fixed Atwood number of $\mathit{A}_T=0.3$: (a) bubble and (b) spike. The seven initial amplitudes in the training dataset, $A_0=k \eta_0$, are distinguished by different colors, and the initial states are marked by open circles. 
}
\label{fig:a0_phase}
\end{figure*}

The bubble and spike branches were treated independently in the SINDy-based identification of their respective dynamical systems.
For both branches, the identified models were formulated within the unified buoyancy--drag structure given by Eq.~\eqref{eq:bd_sindy}.

For the bubble branch, within the prescribed candidate-function library and under the imposed sparsity constraints, the identified buoyancy-correction function and effective drag-coefficient function are given, respectively,
{\color{black}
\begin{align}
B_b(A;A_0,\mathit{A}_T)
&=-0.4194\,A-0.0334\,A\mathit{A}_T-0.0162\,AA_0 \notag\\
&\quad+0.0588\,A^2, \label{eq:B_bubble}\\
D_b(A;A_0,\mathit{A}_T)
&=1.5576\,\mathit{A}_T+0.6890\,A+0.8245\,\mathit{A}_T^2 \notag\\
&\quad-1.3230\,A_0\mathit{A}_T-0.3363\,A\mathit{A}_T-0.0631\,A^2. \label{eq:D_bubble}
\end{align}
}

For the spike branch, the corresponding buoyancy-correction function and effective drag-coefficient function are given, respectively,
{\color{black}
\begin{align}
B_s(A;A_0,\mathit{A}_T)
&=-0.4213\,A+0.2078\,A\mathit{A}_T+0.0574\,A^2 \notag\\
&\quad+0.0844\,A\mathit{A}_T^2-0.0587\,A^2\mathit{A}_T \notag\\
&\quad+0.0002\,A^3, \label{eq:B_spike}\\
D_s(A;A_0,\mathit{A}_T)
&=0.6924-0.5923\,\mathit{A}_T^2-0.6766\,A_0^2 \notag\\
&\quad-0.2379\,A\mathit{A}_T+0.3065\,AA_0+0.0337\,A^2 \notag\\
&\quad+0.7504\,A_0^2\mathit{A}_T+0.3409\,A\mathit{A}_T^2 \notag\\
&\quad-0.3342\,AA_0\mathit{A}_T-0.0582\,A^2\mathit{A}_T \notag\\
&\quad+0.0010\,A^3. \label{eq:D_spike}
\end{align}
}

Here, the subscripts $b$ and $s$ denote the bubble and spike branches, respectively. Substitution of Eqs.~\eqref{eq:B_bubble}--\eqref{eq:D_bubble} or Eqs.~\eqref{eq:B_spike}--\eqref{eq:D_spike} into Eq.~\eqref{eq:bd_sindy} yields the corresponding closed dynamical equations for each branch.

Over the complete training dataset, the mean RMSE of the dimensionless velocity, denoted as $\overline{\mathrm{RMSE}}_U$, were {\color{black}$0.015$} for the bubble branch and {\color{black}$0.027$} for the spike branch. The training dataset
comprised all combinations of
\[
    \mathit{A}_T\in\{0.1,\,0.3,\,0.5,\,0.7,\,0.9\}
\]
and
\[
    \eta_0\in
    \{0.01,\,0.03,\,0.05,\,0.08,\,0.10,\,0.12,\,0.15\}\,\mathrm{m},
\]
resulting in a total of $35$ simulation cases.

For the interpolation test across Atwood numbers, four values not represented
in the training dataset,
$\mathit{A}_T\in\{0.2,\,0.4,\,0.6,\,0.8\}$, were considered, with the
gravitational acceleration and wavenumber held fixed at
$g=9.8\,\mathrm{m\,s^{-2}}$ and $k=2\pi\,\mathrm{m^{-1}}$, respectively.
Combined with the seven initial amplitudes, these parameter combinations
yielded a total of $28$ simulation cases. For brevity, only the cases with
$\eta_0=0.10\,\mathrm{m}$ are presented in Figure~\ref{fig:testat}. 
The $\overline{\mathrm{RMSE}}_U$, averaged over all 28 cases, were {\color{black}$0.013$} and {\color{black}$0.030$} for the bubble and spike branches, respectively. 
These errors remain comparable to their
training-set counterparts, demonstrating that the identified dynamical
equations generalize accurately to unseen Atwood numbers within the
parameter range spanned by the training data.

The results of the generalization tests across gravitational accelerations
are shown in the first and second columns of Figure.~\ref{fig:testgk}.
At a fixed wavenumber of $k=2\pi\,\mathrm{m^{-1}}$, gravitational
accelerations of $g\in\{8,\,10,\,12\}\,\mathrm{m\,s^{-2}}$ were considered. For brevity, only the cases with
$\eta_0=0.1\,\mathrm{m}$ are shown. 
The $\overline{\mathrm{RMSE}}_U$  were {\color{black}$0.014$} and
{\color{black}$0.019$} for the bubble and spike branches, respectively. These very small values indicate that the identified equations extrapolate well in $g$.

The results of the generalization tests across wavenumbers are shown in the
third and fourth columns of Figure~\ref{fig:testgk}. At a fixed gravitational
acceleration of $g=9.8\,\mathrm{m\,s^{-2}}${\color{black}, the two wavenumbers not present in the training data,}
$k\in\{\pi,\,3\pi\}\,\mathrm{m^{-1}}$. {\color{black}The reference panels for
$k=2\pi\,\mathrm{m^{-1}}$, which is the wavenumber of the training set, are shown
for comparison but are excluded from the error statistics.}
Only the cases with $\eta_0=0.1\,\mathrm{m}$ are shown. The corresponding
values of $\overline{\mathrm{RMSE}}_U$ were {\color{black}$0.041$} and
{\color{black}$0.050$} for the
bubble and spike branches, respectively.

The values of $\overline{\mathrm{RMSE}}_U$ for all training and test datasets are summarized in Table~\ref{tab:3model_rmse}, where comparison of trajectory prediction errors for three low-dimensional closure models are shown.
Predicted trajectories are obtained by time integration. 
The RMSE is computed separately for each trajectory, and the reported values are the arithmetic means over all trajectories in each dataset.
The adopted nondimensionalization incorporates the characteristic scales associated with $g$ and $k$, including the velocity scale $\sqrt{g/k}$. 
Across the ranges of gravitational acceleration and wavenumber considered, the results support the conclusion that the identified dynamical equations capture the effective dynamics within a unified dimensionless framework consistent with the underlying similarity scaling, rather than merely providing a case-specific fit tied to a particular combination of dimensional parameters.

\begin{table}[htbp]
\centering
\caption{Comparison of trajectory prediction errors for three low-dimensional closure models. The model columns correspond to the classical buoyancy--drag model, Liu's improved potential-flow theory, and the physics-constrained SINDy model, respectively.}
    \label{tab:3model_rmse}
    \begin{tabular}{lccc}
        \toprule
        & \multicolumn{3}{c}{$\overline{\mathrm{RMSE}}_U$} \\
        \cmidrule(lr){2-4}
        Branch and dataset
        & Buoyancy--drag
        & Liu
        & SINDy \\
        \midrule
        Bubble, training                         & 0.063 & 0.063 & {\color{black}0.015} \\
        Bubble, validation across $\mathit{A}_T$ & 0.064 & 0.063 & {\color{black}0.013} \\
        Bubble, validation across $g$           & 0.059 & 0.037 & {\color{black}0.014} \\
        Bubble, validation across $k$           & {\color{black}0.069} & {\color{black}0.070} & {\color{black}0.041} \\
        Spike, training                          & 0.155 & 0.181 & {\color{black}0.027} \\
        Spike, validation across $\mathit{A}_T$  & 0.140 & 0.163 & {\color{black}0.030} \\
        Spike, validation across $g$            & 0.076 & 0.092 & {\color{black}0.019} \\
        Spike, validation across $k$            & {\color{black}0.132} & 0.140 & {\color{black}0.050} \\
        \bottomrule
    \end{tabular}
\end{table}

\begin{figure*}[htbp]
\centering
\includegraphics[width=\textwidth]{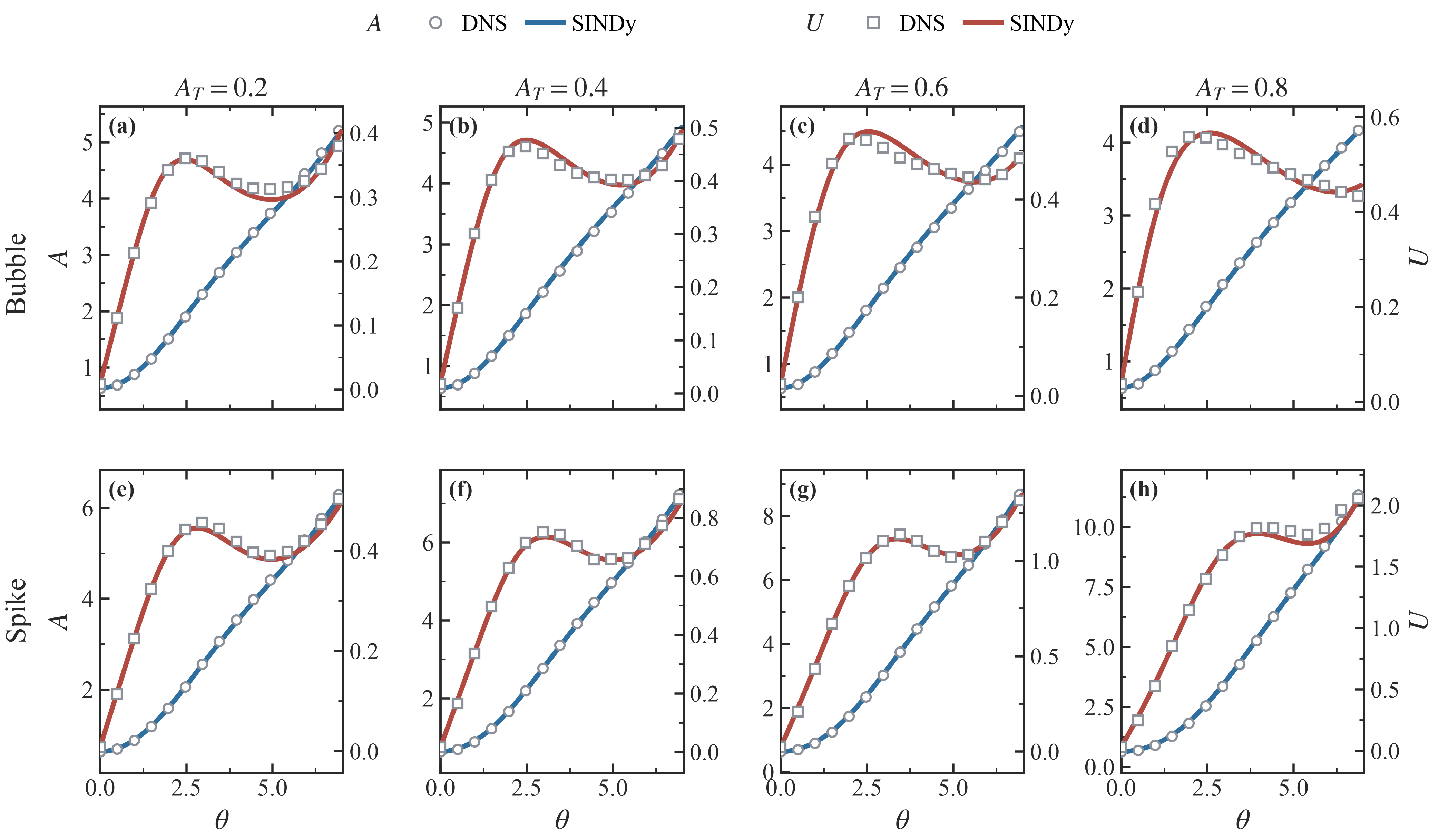}
\caption{Representative interpolation test results for Atwood numbers not included in the training dataset. The first and second rows correspond to the bubble and spike branches, respectively, while the first through fourth columns
correspond to $\mathit{A}_T=0.2$, $0.4$, $0.6$, and $0.8$, respectively. The
initial amplitude is fixed at $\eta_0=0.10\,\mathrm{m}$ for all cases. Open
circles and squares denote the DNS data for the amplitude and velocity,
respectively, whereas the blue and red solid curves represent the
corresponding predictions obtained by forward integration of the
SINDy-identified dynamical equations.}
    \label{fig:testat}
\end{figure*}

To obtain a physically transparent view of how buoyancy and drag compete throughout the evolution and improve the interpretability of the SINDy-identified dynamics, the effective contributions of buoyancy (denoted as $F_{\mathrm{B}}$, see eq.(\ref{eq:force_decomposition})) and drag (denoted as $F_{\mathrm{D}}$,  see also see eq.(\ref{eq:force_decomposition})) to the acceleration predicted by the identified model are shown separately in Figure~\ref{fig:force_Decom}. 
\begin{equation}
    \frac{\mathrm{d}U}{\mathrm{d}\tau}
    =
    \underbrace{
        \mathit{A}_T\,A\left[1+B(A;A_0,\mathit{A}_T)\right]
    }_{F_{\mathrm{B}}}
    +
    \underbrace{
        \left[-D(A;A_0,\mathit{A}_T)U^2\right]
    }_{F_{\mathrm{D}}}.
    \label{eq:force_decomposition}
\end{equation}
In particular, this detailed investigation helps to identify when the amplitude-dependent buoyancy correction $B(A;A_0,\mathit{A}_T)$ versus the velocity-dependent drag term $D(A;A_0,\mathit{A}_T)U^2$ dominates key features such as velocity growth and overshoot.
As the system enters the nonlinear regime, the modification
of the buoyancy term by $B(A;A_0,\mathit{A}_T)$ becomes increasingly pronounced. 
Meanwhile, the drag contribution is jointly determined by the
effective drag coefficient $D(A;A_0,\mathit{A}_T)$ and the squared velocity
$U^2$. The evolving balance between these two contributions governs the net
acceleration and, consequently, the velocity growth, overshoot, and
subsequent dynamical evolution.

\begin{figure*}[htbp]
\centering
\includegraphics[width=\textwidth]{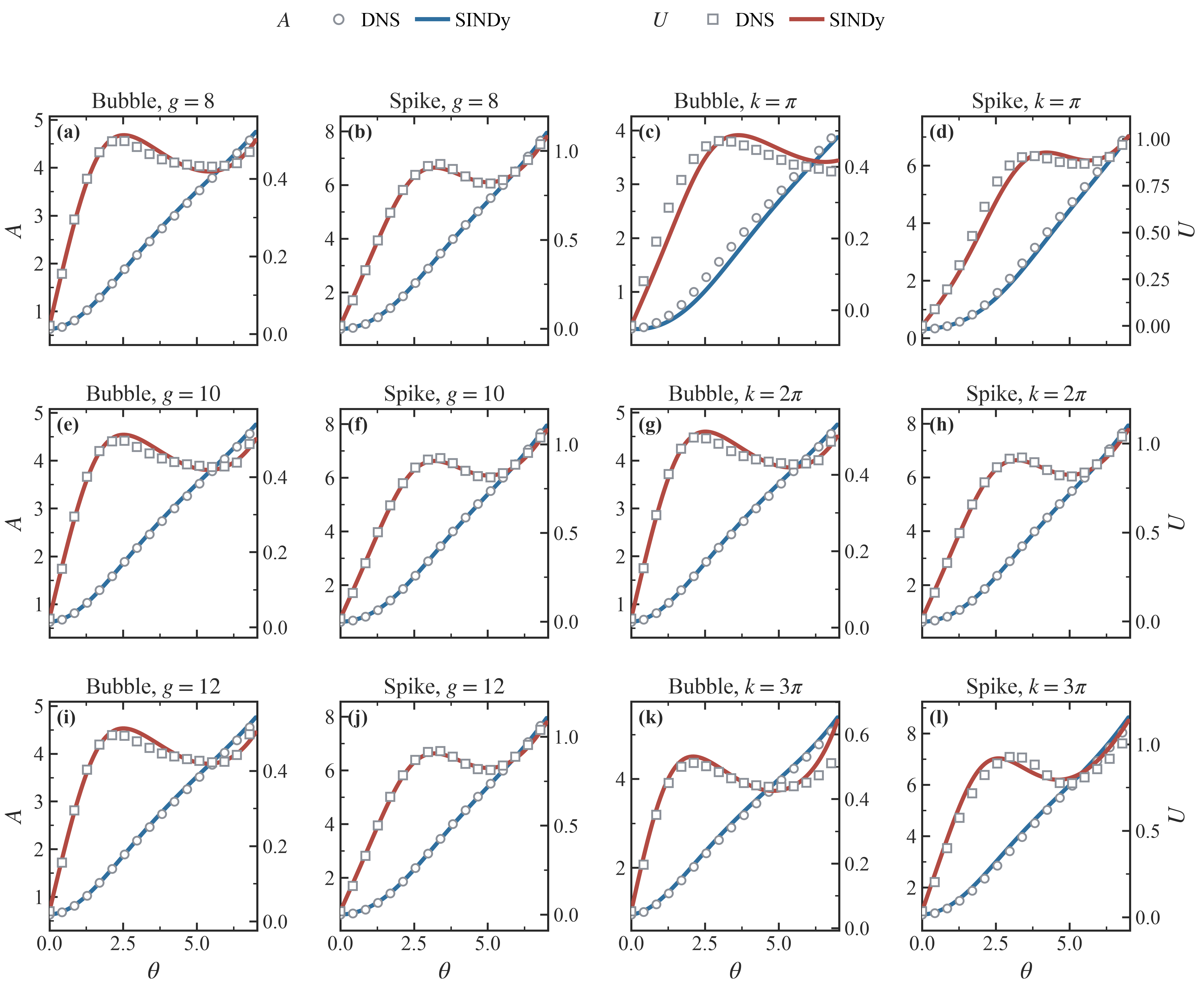}
\caption{Representative generalization tests under wavenumber and gravitational acceleration conditions not included in the training dataset. The first and second columns show the bubble and spike results, respectively, for the generalization test across gravitational accelerations; the first through third rows correspond to $g=8$, $10$, and $12\,\mathrm{m\,s^{-2}}$,
respectively, with $k=2\pi\,\mathrm{m^{-1}}$, $\mathit{A}_T=0.5$, and
$\eta_0=0.10\,\mathrm{m}$. The third and fourth columns show the bubble and
spike results, respectively, for the generalization test across wavenumbers;
the first through third rows correspond to $k=\pi$, $2\pi$, and
$3\pi\,\mathrm{m^{-1}}$, respectively, with
$g=9.8\,\mathrm{m\,s^{-2}}$, $\mathit{A}_T=0.5$, and
$\eta_0=0.10\,\mathrm{m}$;  Open circles and squares denote the DNS data for the
amplitude and velocity, respectively, whereas the blue and red solid curves
represent the corresponding predictions obtained by forward integration of
the SINDy-identified dynamical equations.}
\label{fig:testgk}
\end{figure*}

\begin{figure*}[htbp]
\centering
\includegraphics[width=\textwidth]{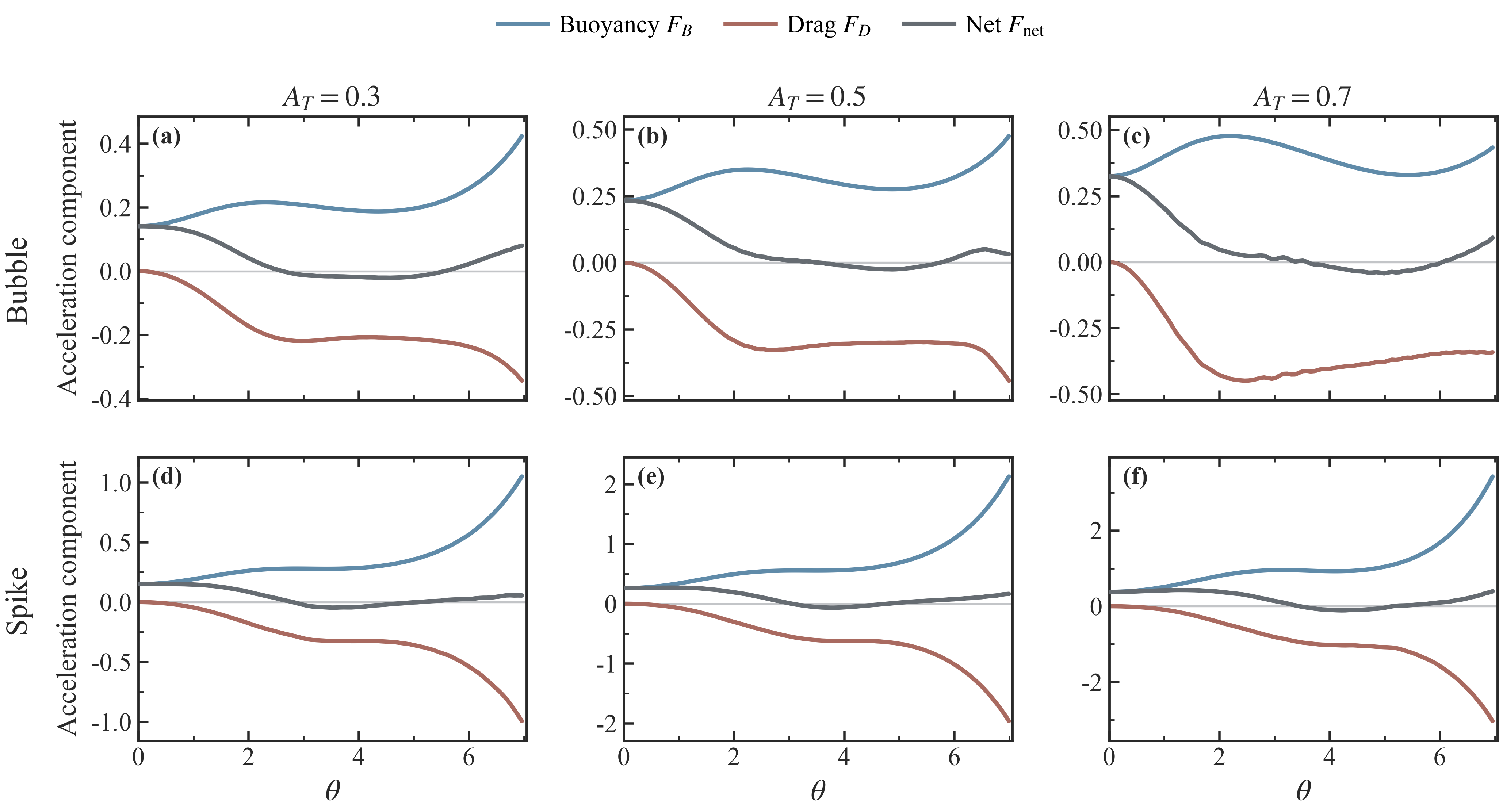}
\caption{Buoyancy--drag decomposition and the resulting net acceleration of the
SINDy-identified dynamical equations for
$\mathit{A}_T\in\{0.3,\,0.5,\,0.7\}$. The upper and lower rows correspond to
the bubble and spike branches, respectively, with the initial amplitude
fixed at $\eta_0=0.10\,\mathrm{m}$ in all cases. Shown separately are the
effective buoyancy contribution,
$F_{\mathrm{B}}=\mathit{A}_T\,A
[1+B(A;A_0,\mathit{A}_T)]$, the effective drag contribution,
$F_{\mathrm{D}}=-D(A;A_0,\mathit{A}_T)U^2$, and the resulting net
acceleration,
$F_{\mathrm{net}}=F_{\mathrm{B}}+F_{\mathrm{D}}
=\mathrm{d}U/\mathrm{d}\tau$.}
\label{fig:force_Decom}
\end{figure*}

As indicated by the DNS results presented in Figure~\ref{fig:a0_phase}, the peak bubble-tip velocity shows a strong dependence on the initial amplitude $\eta_0$, with an overall tendency to increase as $\eta_0$ grows.

For the initial interface profile $\eta(x,0)=-\eta_0\cos(kx)$, with $\eta_0>0$, the bubble tip is located at $x=\pm\lambda/2$. The signed curvature at the bubble tip is given by
\begin{equation}
\xi_0 = \left.  \frac{\partial^2\eta/\partial x^2}
{\left[  1+\left(\partial\eta/\partial x\right)^2  \right]^{3/2}}
\right|_{x=\pm\lambda/2}  =   -k^2\eta_0,
\label{eq:initial_curvature}
\end{equation}
where the final equality follows directly from
\[
\left.\frac{\partial\eta}{\partial x}\right|_{x=\pm\lambda/2}=0,
\qquad
\left.\frac{\partial^2\eta}{\partial x^2}\right|_{x=\pm\lambda/2}
=-k^2\eta_0.
\]

Therefore, for a fixed wavenumber $k$, with all other parameters kept
constant, the magnitude of the initial curvature increases with the
initial amplitude according to $|\xi_0|=k^2\eta_0$. In single-mode RTI, the initial amplitude therefore
encodes not only the displacement scale of the initial perturbation but also
the associated initial interfacial geometry. This geometric dependence can
modify the subsequent interface evolution, thereby influencing the
effective buoyancy--drag balance and the resulting velocity overshoot.

During the early linear regime, when the velocity remains small, the
contribution of velocity-dependent drag to the acceleration is negligible.
Consequently, gravitational driving arising from the density contrast
dominates, maintaining a positive bubble acceleration. 
The bubble velocity therefore increases gradually from zero, while the interfacial perturbation subsequently exhibits approximately exponential growth. 
In the vicinity of the perturbed interface, the density and pressure
gradients are not parallel, providing a source of vorticity through the
baroclinic torque. As shown in Figure~\ref{fig:vorticity}, oppositely signed
vorticity layers form on the two sides of the bubble and spike interfaces.
The associated vorticity-induced velocity field contributes to the transverse
deformation of the interface.
For the bubble, the vorticity-induced transverse flows on either side
of the tip contribute to its lateral broadening and thereby influence the
evolution of the bubble-tip curvature. 
During the early nonlinear regime, the bubble tip continues to advance upward, accompanied by a progressive increase in the magnitude of its curvature. As the tip subsequently broadens laterally, the curvature magnitude reaches a maximum and then gradually relaxes toward a quasi-steady value. This nonmonotonic evolution reflects the transition from initial tip sharpening to subsequent lateral broadening.
Figure~\ref{fig:cur} illustrates this nonmonotonic evolution. Across the
parameter ranges considered, the normalized-curvature histories exhibit only
modest variations for different values of $\mathit{A}_T$ at fixed $A_0$,
whereas substantially different extrema and relaxation behaviors are observed
for different values of $A_0$ at fixed $\mathit{A}_T$. These results suggest
that, within the parameter range considered, variations in $A_0$ exert a
stronger influence on the normalized-curvature evolution than variations in
$\mathit{A}_T$.

The temporal evolution of the dimensionless velocity is governed by the
dynamic balance between the effective buoyancy and drag contributions, see eq. (\ref{eq:force_decomposition}). When $F_{\mathrm{B}}>\lvert F_{\mathrm{D}}\rvert$, the net acceleration is positive, and the bubble-tip velocity continues to increase. When $F_{\mathrm{B}}=\lvert F_{\mathrm{D}}\rvert$, the net acceleration vanishes,
$\mathrm{d}U/\mathrm{d}\tau=0$. When the net acceleration subsequently changes
sign from positive to negative, the velocity attains a local maximum,
corresponding to the velocity overshoot. When
$F_{\mathrm{B}}<\lvert F_{\mathrm{D}}\rvert$, the net acceleration is negative,
and the bubble-tip velocity enters a decelerating regime and subsequently
relaxes toward a lower or quasi-steady value.

At a fixed wavenumber, a larger initial amplitude corresponds to a greater
magnitude of the initial curvature and may thereby modify the subsequent
evolution of the interfacial geometry, ultimately influencing the dynamic
balance between buoyancy and drag and the resulting velocity overshoot.
Within the parameter range considered, this modulation increases the
peak velocity. The integral-form SINDy framework, constrained by the
buoyancy--drag structure, extends the classical buoyancy--drag model by
introducing effective closure functions that depend on the instantaneous
amplitude $A$, the initial amplitude $A_0$, and the Atwood number
$\mathit{A}_T$. The resulting model is therefore capable of reproducing both
the velocity overshoot and the subsequent relaxation process. At later
stages of the evolution, the effective buoyancy contribution increases and
again exceeds the magnitude of the drag contribution, causing the net
acceleration to become positive once more. Consequently, the identified
model also captures the dynamical behavior associated with the
reacceleration regime.

\begin{figure*}[htbp]
\centering
\includegraphics[width=\textwidth]{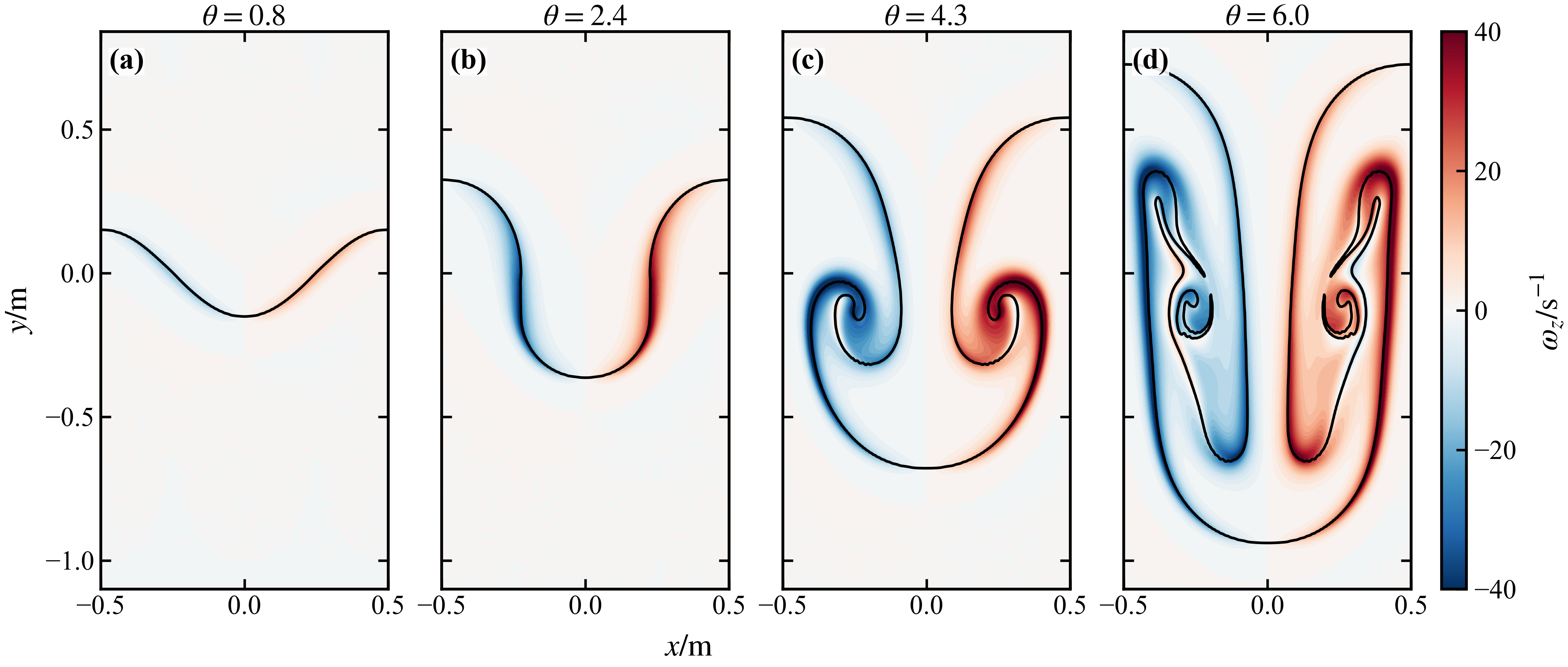}
\caption{Vorticity fields and interfacial morphologies at four instants
for $\mathit{A}_T=0.3$ and $\eta_0=0.10\,\mathrm{m}$: (a) $\theta=t\sqrt{\mathit{A}_T\,gk}=0.8$; (b) $\theta=t\sqrt{\mathit{A}_T\,gk}=2.4$; (c) $\theta=t\sqrt{\mathit{A}_T\,gk}=4.3$; and (d) $\theta=t\sqrt{\mathit{A}_T\, gk}=6.0$.
The black curves denote the interface, while red and blue indicate positive and negative vorticity, respectively. }
\label{fig:vorticity}
\end{figure*}

\begin{figure*}[htbp]
\centering
\includegraphics[width=\textwidth]{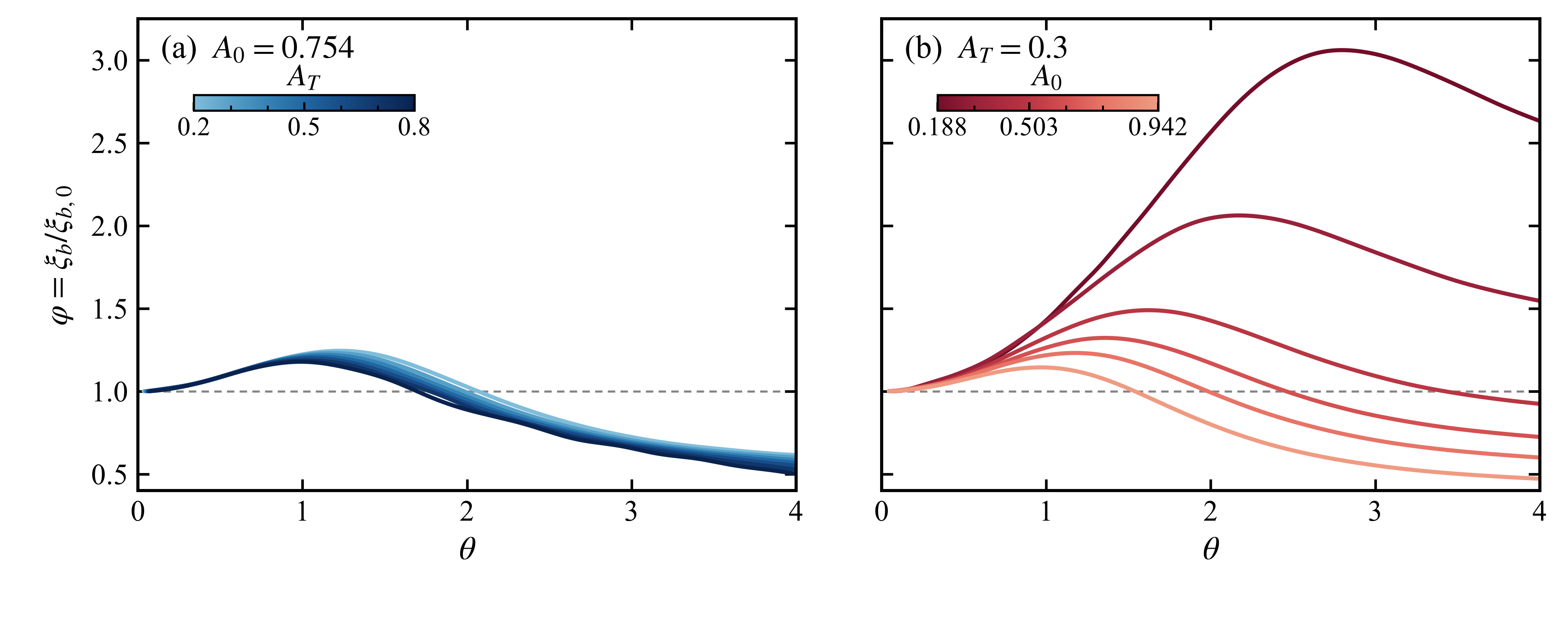}
\caption{Evolution of the normalized bubble-tip curvature,
$\varphi(\tau)=\xi(\tau)/\xi_0$, where $\xi(\tau)$ denotes the bubble-tip
curvature and $\xi_0=-k^2\eta_0$ is the initial bubble-tip curvature.
Panel (a) shows the evolution for different $\mathit{A}_T$ at fixed
$A_0=0.754$, whereas panel (b) shows the evolution for different $A_0$ at
fixed $\mathit{A}_T=0.3$. Following the attainment of an extremum, the
normalized curvature gradually relaxes toward a quasi-steady value,
reflecting the transition from initial tip sharpening to subsequent lateral
broadening of the bubble tip.}
    \label{fig:cur}
\end{figure*}

%==================================================================
\section{Conclusion}
%==================================================================
\label{sec:conclusion}
We have developed a physics-constrained, data-driven framework for
identifying low-dimensional dynamical equations for the bubble and spike
branches from time-resolved data. The framework combines integral-form
SINDy with a common buoyancy--drag structure applied separately to the two
branches. This formulation enables the sparse identification of
branch-specific effective closure functions that explicitly depend on the
instantaneous amplitude $A$, the initial amplitude $A_0$, and the Atwood
number $\mathit{A}_T$. By formulating the regression problem in terms of
finite-interval integrals, the method avoids direct numerical differentiation
of the velocity time series and thereby mitigates noise amplification in
acceleration estimation.

Over the parameter range considered for single-mode RTI, the identified models accurately reproduce the temporal evolution of the bubble and spike amplitudes and velocities, including the velocity overshoot--relaxation behavior and the subsequent late-time reacceleration. The models also retain high predictive accuracy at Atwood numbers, wavenumbers, and gravitational accelerations not included in the training set. These results demonstrate that the identified formulation generalizes across the considered parameter variations under the adopted nondimensionalization. The explicit dependence on the initial perturbation amplitude enables the models to reproduce the variations in peak velocity and subsequent velocity evolution among different initial conditions.

The buoyancy--drag decomposition shows that velocity overshoot is closely related to the competition between effective buoyancy and drag. A change in net acceleration from positive to negative corresponds to the velocity peak, whereas a return to positive net acceleration at later times corresponds to reacceleration. The initial amplitude carries information about the interface geometry. Combining this information with analyses of phase trajectories and tip curvature helps interpret the dynamical differences across initial conditions. The identified state-dependent closures provide a compact dynamical description of these differences, but their links to curvature and vorticity require further study. Future work will explicitly incorporate variables such as curvature and vorticity into the state space to further assess and refine the physical interpretation of the closure model. The framework will also be extended to RTI and RMI with multimode perturbations and under conditions involving viscosity, compressibility, interfacial tension, and variable acceleration.

\acknowledgements
This work is supported by the Science Challenge Project (No. TZ2025012).

\appendix

\section{Numerical Simulation Setup}
\label{app:openfoam}

The numerical simulations are performed using the \texttt{interFoam} solver in OpenFOAM v8. The computational domain is defined by $(x,y)\in[-L_x,L_x]\times[-L_y,L_y]$. The initial interface profile is prescribed as a single-mode sinusoidal perturbation, $y_{\mathrm{interface}}(x,0)= -\eta_0 \cos\left(2\pi x/\lambda\right)$, where $\eta_0$ and $\lambda$ denote the initial perturbation amplitude and wavelength, respectively. For simulations with different wavenumbers $k=2\pi/\lambda$, the computational domain in the transverse direction spans one full wavelength, such that $2L_x=\lambda$. The main physical parameters and numerical settings are summarized in Table~\ref{tab:openfoam}.

\begin{table}[htbp]
\centering
\caption{Parameter settings for the OpenFOAM simulations}
\label{tab:openfoam}
\begin{tabular}{ll}
\hline\hline
Parameter & Setting \\
\hline
Dynamic viscosity
    & $\mu_1=\mu_2=0.003\, \mathrm{Pa\cdot s}$ (approximately inviscid) \\
Surface tension
    & $\sigma=0$ (no interfacial tension) \\
Gravitational acceleration
    & $g=9.8\,\mathrm{m\,s^{-2}}$ (directed downward) \\
Left and right boundaries
    & Periodic boundary conditions \\
Top and bottom boundaries (phase fraction)
    & \texttt{zeroGradient} \\
Top and bottom boundaries (velocity)
    & \texttt{noSlip} \\
Convection-term discretization
    & \texttt{Gauss vanLeer} (phase fraction) \\
Gradient discretization
    & \texttt{Gauss linear} (second-order accurate) \\
Time integration
    & First-order implicit Euler scheme \\
Baseline mesh
    & $N_x \times N_y = 256 \times 512$ \\
\hline\hline
\end{tabular}
\end{table}

To verify that the baseline mesh provides sufficient spatial resolution, a grid-convergence study was performed using three meshes with resolutions of $256\times512$, $512\times1024$, and $800\times1600$. Figure~\ref{fig:convergence} shows the temporal evolution of the bubble--spike separation. The results demonstrate that the predictions obtained on the $256\times512$ mesh agree closely with those obtained on the two finer meshes. This resolution is therefore sufficient to resolve the principal features of the nonlinear evolution of the RTI and satisfies the spatial-resolution requirements of the present numerical simulations.

\begin{figure}[htbp]
\centering
\includegraphics[width=0.45\textwidth]{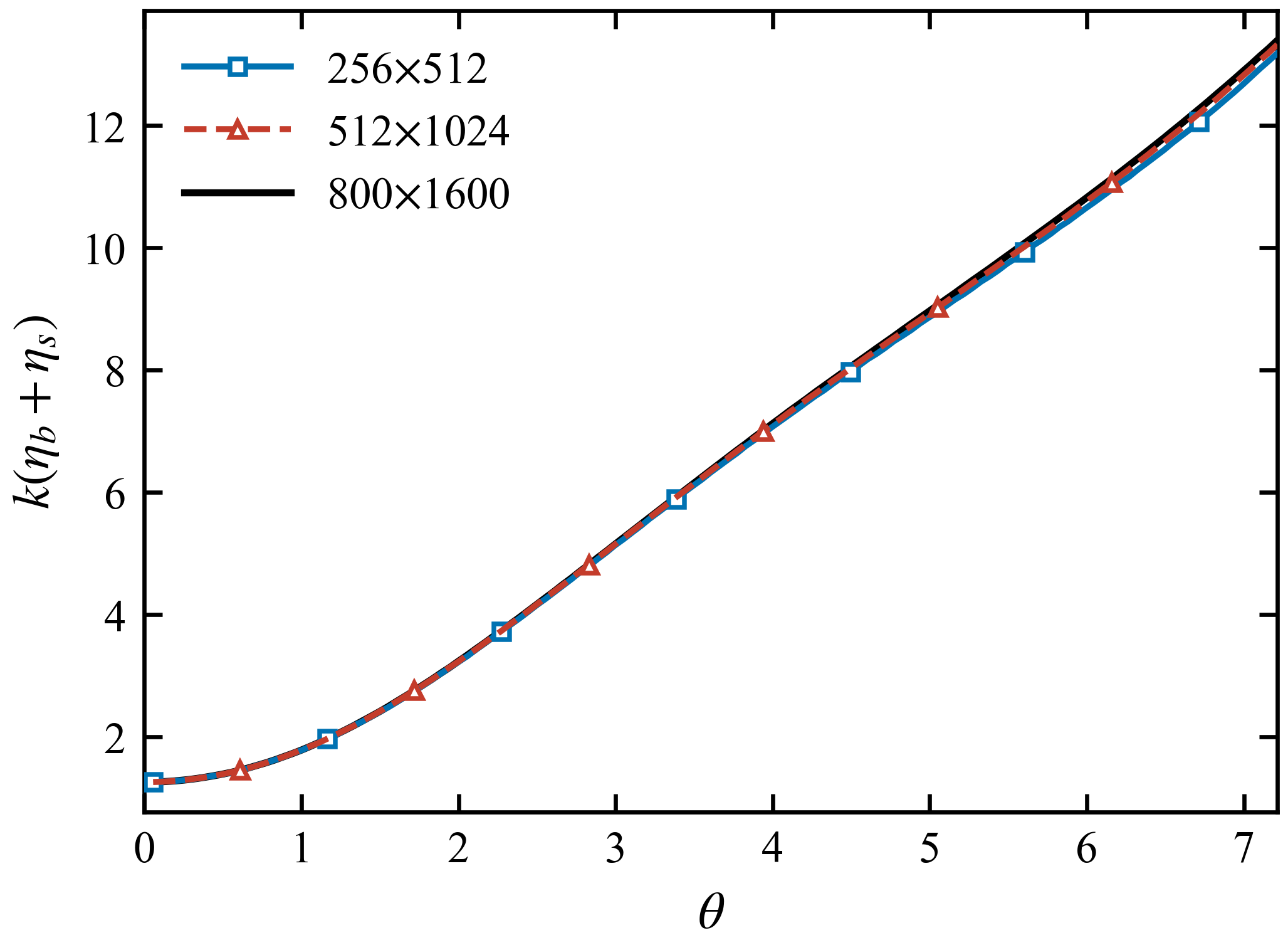}
\caption{Temporal evolution of the bubble--spike distance in the grid-convergence study. The blue squares, red triangles, and black solid line represent the numerical results obtained on the $256\times512$, $512\times1024$, and $800\times1600$ meshes, respectively.}
\label{fig:convergence}
\end{figure}

\section{Sensitivity Analysis of Integral-Form SINDy}
\label{app:sensitivity}

This appendix examines the sensitivity of the governing equations identified by integral-form SINDy to the sparsification threshold and the choice of candidate-function library. In all tests, the buoyancy--drag structure used in the main text is held fixed.

\subsection*{Selection of the Sparsification Threshold}
\label{app:threshold}

The sparsification threshold $\delta_I$ is selected by scanning the interval from $0$ to $3.0$ in increments of $0.1$, i.e., $\delta_I\in\{0,0.1,0.2,\ldots,3.0\}$. For each candidate threshold, the score $\mathcal{J}=\mathrm{RMSE}_{\mathrm{train}}(U)+\lambda_s N_{\mathrm{active}}$ is evaluated according to Eq.~\eqref{eq:threshold_score} to balance the training error against model complexity. Here, $\lambda_s=2\times10^{-3}$, and $N_{\mathrm{active}}$ denotes the number of retained candidate terms. Table~\ref{tab:threshold_scan} summarizes representative results from scans of the sparsification threshold, using a second-order candidate-function library for the bubble ($p={\color{black}17}$) and a third-order candidate-function library for the spike ($p={\color{black}36}$). According to the scoring criterion defined above, the selected bubble model is obtained at $\delta_I=0.3$ and contains ${\color{black}10}$ active terms, whereas the selected spike model is obtained at $\delta_I=2.5$ and contains ${\color{black}17}$ active terms. The corresponding entries are highlighted in boldface in the table.

\begin{table*}[htbp]
\centering
\caption{
Representative results from the normalized-coefficient threshold scan. 
The threshold $\delta_I$ is applied to the indicator $I_m=\lvert\widetilde{c}_m\rvert$
formed from the rescaled coefficient vector of Eq.~\eqref{eq:normalized_coefficients}. 
The score is $\mathcal{J}=\mathrm{RMSE}_{\mathrm{train}}(U)+0.002N_{\mathrm{active}}$, where $N_{\mathrm{active}}$ denotes the number of retained candidate terms. The bubble and spike use second-order and third-order candidate-function libraries containing ${\color{black}17}$ and ${\color{black}36}$ terms, respectively. Boldface indicates the selected model for each branch.
}
\label{tab:threshold_scan}
\scriptsize
\setlength{\tabcolsep}{4.5pt}
\begin{tabular}{ccccccc}
\hline\hline
& \multicolumn{3}{c}{Bubble (second-order library)}
& \multicolumn{3}{c}{Spike (third-order library)}\\
\cline{2-4}\cline{5-7}
$\delta_I$
& $N_{\mathrm{active}}$
& $\mathrm{RMSE}_{\mathrm{train}}(U)$
& $\mathcal{J}$
& $N_{\mathrm{active}}$
& $\mathrm{RMSE}_{\mathrm{train}}(U)$
& $\mathcal{J}$\\
\hline
{\color{black}0.0} & {\color{black}17} & {\color{black}0.0147} & {\color{black}0.0487} & {\color{black}36} & {\color{black}0.0193} & {\color{black}0.0913}\\
{\color{black}0.1} & {\color{black}13} & {\color{black}0.0146} & {\color{black}0.0406} & {\color{black}35} & {\color{black}0.0193} & {\color{black}0.0893}\\
{\color{black}0.2} & {\color{black}11} & {\color{black}0.0144} & {\color{black}0.0364} & {\color{black}35} & {\color{black}0.0193} & {\color{black}0.0893}\\
{\color{black}0.3} & \textbf{\color{black}10} & \textbf{\color{black}0.0145} & \textbf{\color{black}0.0345}
& {\color{black}34} & {\color{black}0.0193} & {\color{black}0.0873}\\
{\color{black}0.4} & {\color{black}6} & {\color{black}0.0235} & {\color{black}0.0355} & {\color{black}33} & {\color{black}0.0195} & {\color{black}0.0855}\\
{\color{black}0.5} & {\color{black}8}  & {\color{black}0.0191} & {\color{black}0.0351} & {\color{black}33} & {\color{black}0.0195} & {\color{black}0.0855}\\
{\color{black}0.6} & {\color{black}6}  & {\color{black}0.0235} & {\color{black}0.0355} & {\color{black}31} & {\color{black}0.0199} & {\color{black}0.0819}\\
{\color{black}0.7} & {\color{black}---} & {\color{black}---} & {\color{black}---} & {\color{black}31} & {\color{black}0.0199} & {\color{black}0.0819}\\
{\color{black}0.8} & {\color{black}---} & {\color{black}---} & {\color{black}---} & {\color{black}28} & {\color{black}0.0198} & {\color{black}0.0758}\\
{\color{black}0.9} & {\color{black}---} & {\color{black}---} & {\color{black}---} & {\color{black}26} & {\color{black}0.0195} & {\color{black}0.0715}\\
{\color{black}1.0} & {\color{black}3}  & {\color{black}0.0498} & {\color{black}0.0558} & {\color{black}26} & {\color{black}0.0201} & {\color{black}0.0721}\\
{\color{black}1.1} & {\color{black}3}  & {\color{black}0.0747} & {\color{black}0.0807} & {\color{black}26} & {\color{black}0.0201} & {\color{black}0.0721}\\
{\color{black}1.6} & {\color{black}3}  & {\color{black}0.0498} & {\color{black}0.0558} & {\color{black}19} & {\color{black}0.0270} & {\color{black}0.0650}\\
{\color{black}2.1} & {\color{black}3}  & {\color{black}0.0498} & {\color{black}0.0558} & {\color{black}19} & {\color{black}0.0270} & {\color{black}0.0650}\\
{\color{black}2.5} & {\color{black}3}  & {\color{black}0.0498} & {\color{black}0.0558}
& \textbf{\color{black}17} & \textbf{\color{black}0.0271} & \textbf{\color{black}0.0611}\\
{\color{black}2.7} & {\color{black}3}  & {\color{black}0.0498} & {\color{black}0.0558} & {\color{black}17} & {\color{black}0.0271} & {\color{black}0.0611}\\
{\color{black}3.0} & {\color{black}3}  & {\color{black}0.0498} & {\color{black}0.0558} & {\color{black}17} & {\color{black}0.0271} & {\color{black}0.0611}\\
\hline\hline
\end{tabular}
\end{table*}

\subsection*{Selection of the Candidate-Function Library}
\label{app:function}

To isolate the effect of the candidate-function library on the identified equations, we keep fixed the buoyancy--drag structure in Eq.~\eqref{eq:bd_sindy}, the training/test split, the integral-form regression procedure, the normalized coefficient-threshold grid, and the model-selection criterion defined in Eq.~\eqref{eq:threshold_score}. For each library, the threshold scan and coefficient identification are performed independently, and the resulting model is evaluated on the train set through forward integration.

For a given polynomial order, the candidate library is constructed from all
monomials in the three variables $A$, $A_0$, and $\mathit{A}_T$. Here, $A=k\eta$ and
$A_0=k\eta_0$ denote the nondimensional amplitude and nondimensional initial
amplitude, respectively, while $\mathit{A}_T$ denotes the Atwood number. The
order-$n$ candidate library $\mathcal{L}_n$ is defined by
{\color{black}
\begin{equation}
\mathcal{L}_n:\quad
\begin{cases}
B \in \mathrm{span} \left\{
A^i A_0^{\,j} \mathit{A}_T^{\ell}
\,\middle|\,
i,j,\ell\in\mathbb{Z}_{\geq 0},\;
1\leq i+j+\ell\leq n,\; i+j\geq 1
\right\},\\[4pt]
D \in \mathrm{span} \left(
\{1\}\cup  \left\{
A^i A_0^{\,j} \mathit{A}_T^{\ell}
\,\middle|\,
i,j,\ell\in\mathbb{Z}_{\geq 0},\;
1\leq i+j+\ell\leq n
\right\}
\right).
\end{cases}
\label{eq:library_definition_2}
\end{equation}
}
{\color{black}
The buoyancy library excludes the monomials of the form $\mathit{A}_T^{\ell}$ with $\ell\geq1$; these are precisely the terms that survive the limit $A\to0$, $A_0\to0$. Their removal enforces $B(0;0,\mathit{A}_T)=0$ exactly, so that Eq.~\eqref{eq:bd_sindy} reduces identically to the classical linear RTI equation. The drag library is retained in full, since the drag contribution $D U^2$ is $O(U^2)$ and does not enter the linear limit.
}

\begin{table}[htbp]
\centering
\caption{Number of terms in the candidate-function libraries of different polynomial orders. Here, $N_B$ and $N_D$ denote the numbers of terms in the buoyancy and drag libraries, respectively, with $N_D$ including the constant term; $N_{\mathrm{total}}=N_B+N_D$.}
\label{tab:library_terms}
\small
\setlength{\tabcolsep}{6pt}
\begin{tabular}{ccccc}
\hline\hline
Polynomial order $n$ & Candidate library & $N_B$ & $N_D$ & $N_{\mathrm{total}}$\\
\hline
1 & $\mathcal{L}_1$ & {\color{black}2}  & 4  & {\color{black}6}\\
2 & $\mathcal{L}_2$ & {\color{black}7}  & 10 & {\color{black}17}\\
3 & $\mathcal{L}_3$ & {\color{black}16} & 20 & {\color{black}36}\\
4 & $\mathcal{L}_4$ & {\color{black}30} & 35 & {\color{black}65}\\
5 & $\mathcal{L}_5$ & {\color{black}50} & 56 & {\color{black}106}\\
\hline\hline
\end{tabular}
\end{table}

Table~\ref{tab:library_terms} summarizes the number of candidate terms in the library at each polynomial order. Table~\ref{tab:library_comparison} summarizes the forward-integration
prediction errors for candidate-function libraries of different polynomial
orders after sparsification. The training set contains $35$ trajectories
with $\mathit{A}_T\in\{0.1,0.3,0.5,0.7,0.9\}$, whereas the test set contains $28$
trajectories at the held-out values
$\mathit{A}_T\in\{0.2,0.4,0.6,0.8\}$. For each dataset, the reported error is the
arithmetic mean of the trajectory-wise $\mathrm{RMSE}(U)$ values. 

The optimal candidate-function library is selected by minimizing an objective
function of the same form as that in Eq.~\eqref{eq:threshold_score}. When
comparing different libraries, the training-set $\mathrm{RMSE}(U)$ is used as
the error term, while the complexity penalty accounts for model complexity:
\begin{equation}
\mathcal{J}_{\mathrm{lib}}
=
\mathrm{RMSE}_{\mathrm{train}}(U)
+\lambda_s N_{\mathrm{active}},
\label{eq:library_score}
\end{equation}
Here, $\mathrm{RMSE}_{\mathrm{train}}(U)$ denotes the velocity-reconstruction
RMSE evaluated on the train set ,
$N_{\mathrm{active}}$ is the number of terms retained after threshold-based
sparsification, and $\lambda_s=2\times10^{-3}$. By penalizing both the train
error and model complexity, this objective discourages the selection of
overly complex high-order models that overfit the data.

\begin{table*}[htbp]
\centering
\caption{Forward-integration prediction errors of models identified using candidate-function libraries of orders $1$--$5$. Here, $N_{\mathrm{active}}$ denotes the number of terms retained after threshold-based sparsification. The training and test
$\mathrm{RMSE}(U)$ values are arithmetic means over individual trajectories
in the training set and the test set at held-out $\mathit{A}_T$ values, respectively.
$\mathcal{J}_{\mathrm{lib}}$ is the library-selection objective defined in
Eq.~\eqref{eq:library_score}. {\color{black}On the buoyancy-constrained library,
no stable first-order spike model is found, so that entry is left blank.} Boldface identifies the library selected
for each branch: $\mathcal{L}_2$ for the bubble and $\mathcal{L}_3$ for the
spike.}
\label{tab:library_comparison}
\scriptsize
\setlength{\tabcolsep}{4pt}
\begin{tabular}{lcccccccc}
\hline\hline
& \multicolumn{4}{c}{Bubble} & \multicolumn{4}{c}{Spike}\\
\cline{2-5}\cline{6-9}
Candidate library
& $N_{\mathrm{active}}$
& $\mathrm{RMSE}_{\mathrm{train}}(U)$
& $\mathrm{RMSE}_{\mathrm{test}}(U)$
& $\mathcal{J}_{\mathrm{lib}}$
& $N_{\mathrm{active}}$
& $\mathrm{RMSE}_{\mathrm{train}}(U)$
& $\mathrm{RMSE}_{\mathrm{test}}(U)$
& $\mathcal{J}_{\mathrm{lib}}$\\
\hline
$\mathcal{L}_1$ & {\color{black}3}  & {\color{black}0.0483} & {\color{black}0.0465} & {\color{black}0.0543} & {\color{black}---} & {\color{black}---} & {\color{black}---} & {\color{black}---}\\
$\mathcal{L}_2$ & \textbf{\color{black}10} & \textbf{\color{black}0.0145} & \textbf{\color{black}0.0134} & \textbf{\color{black}0.0345} & {\color{black}11} & {\color{black}0.1210} & {\color{black}0.1202} & {\color{black}0.1430}\\
$\mathcal{L}_3$ & {\color{black}7} & {\color{black}0.0248} & {\color{black}0.0241} & {\color{black}0.0388} & \textbf{\color{black}17} & \textbf{\color{black}0.0271} & \textbf{\color{black}0.0296} & \textbf{\color{black}0.0611}\\
$\mathcal{L}_4$ & {\color{black}11} & {\color{black}0.0218} & {\color{black}0.0212} & {\color{black}0.0438} & {\color{black}23} & {\color{black}0.0204} & {\color{black}0.0177} & {\color{black}0.0664}\\
$\mathcal{L}_5$ & {\color{black}10} & {\color{black}0.0215} & {\color{black}0.0217} & {\color{black}0.0415} & {\color{black}79} & {\color{black}0.0148} & {\color{black}0.0136} & {\color{black}0.1728}\\
\hline\hline
\end{tabular}
\end{table*}

Balancing the train error against model complexity, $\mathcal{L}_2$ yields the lowest $\mathcal{J}_{\mathrm{lib}}$ for the bubble branch (${\color{black}0.0345}$), whereas $\mathcal{L}_3$ yields the lowest value for the spike branch (${\color{black}0.0611}$). Accordingly, the models identified using $\mathcal{L}_2$ and $\mathcal{L}_3$ are selected for the bubble and spike branches, respectively. They retain ${\color{black}10}$ and ${\color{black}17}$ active terms, with corresponding test $\mathrm{RMSE}$ values of ${\color{black}0.0134}$ and ${\color{black}0.0296}$.

{\color{black}
\section{Necessity of the Initial Amplitude: Ablation Study}

\label{app:a0_ablation}

The main text argues, from the intersecting phase-space trajectories of
Fig.~\ref{fig:a0_phase}, that $(A,U;\mathit{A}_T)$ does not uniquely determine
the subsequent dynamics, so that $A_0$ must enter the identified equations
explicitly. To quantify the cost of omitting it, we repeat the identification
of Appendix~\ref{app:function} with every $A_0$-containing monomial ($j\geq1$)
deleted from the library $\mathcal{L}_n$ of
Eq.~\eqref{eq:library_definition_2}, in $B$ and in $D$ alike, so that the
ablated model depends on $(A,\mathit{A}_T)$ only. The two libraries are compared \emph{best versus best}: each is scanned over
orders $1$--$5$ and its own optimum under
$\mathcal{J}=\mathrm{RMSE}_{\mathrm{train}}(U)+2\times10^{-3}N_{\mathrm{active}}$
is retained, with no test-set information entering the selection.

\begin{table}[htbp]
\color{black}
\centering
\caption{Effect of removing $A_0$ from the candidate-function library. Here $n$ is the polynomial order, $N_{\mathrm{lib}}$ the library size, $N_{\mathrm{active}}$ the number of retained terms, and $\mathrm{RMSE}_{\mathrm{test}}(U)$ the error over the $28$ test trajectories at held-out $\mathit{A}_T$. Boldface marks the better model of each branch.}
\label{tab:a0_ablation}
\small
\setlength{\tabcolsep}{5pt}
\begin{tabular}{lccccc}
\hline\hline
Branch & Library & $n$ & $N_{\mathrm{lib}}$
& $N_{\mathrm{active}}$ & $\mathrm{RMSE}_{\mathrm{test}}(U)$\\
\hline
\multirow{2}{*}{Bubble}
 & with $A_0$    & 2 & 17 & 10 & \textbf{0.0134}\\
 & without $A_0$ & 2 & 9  & 6  & 0.0224\\
\hline
\multirow{2}{*}{Spike}
 & with $A_0$    & 3 & 36 & 17 & \textbf{0.0296}\\
 & without $A_0$ & 3 & 16 & 12 & 0.0540\\
\hline\hline
\end{tabular}
\end{table}

Removing $A_0$ raises the test error from $0.0134$ to $0.0224$ for the bubble and from $0.0296$ to $0.0540$ for the spike. The initial amplitude
is therefore not a redundant descriptor that a richer polynomial library could replace.
}
\bibliography{ref.bib} 

@article{Rayleigh,
author = {Rayleigh},
title = {Investigation of the Character of the Equilibrium of an Incompressible Heavy Fluid of Variable Density},
journal = {Proceedings of the London Mathematical Society},
volume = {s1-14},
number = {1},
pages = {170-177},
doi = {https://doi.org/10.1112/plms/s1-14.1.170},
url = {https://londmathsoc.onlinelibrary.wiley.com/doi/abs/10.1112/plms/s1-14.1.170},
eprint = {https://londmathsoc.onlinelibrary.wiley.com/doi/pdf/10.1112/plms/s1-14.1.170},
year = {1882}
}

@article{Taylor1950instability,
  title={The instability of liquid surfaces when accelerated in a direction perpendicular to their planes. I},
  author={Taylor, Geoffrey Ingram},
  journal={Proceedings of the Royal Society of London. Series A. Mathematical and Physical Sciences},
  volume={201},
  number={1065},
  pages={192--196},
  year={1950},
  publisher={The Royal Society London}
}

@article{2014NIFReview,
  title={Review of the National Ignition Campaign 2009-2012},
  author={ Lindl, John  and  Landen, Otto  and  Edwards, John  and  Moses, Ed },
  journal={Physics of Plasmas},
  volume={21},
  number={2},
  pages={020501},
  year={2014},
  doi={10.1063/1.4865400}
}

@article{2016Inertial,
  title={Inertial-Confinement fusion with lasers},
  author={ Betti, R.  and  Hurricane, O. A. },
  journal={Nature Physics},
  volume={12},
  number={5},
  pages={435-448},
  year={2016},
}

@article{cabot2006supernovae,
  title={Reynolds number effects on Rayleigh--Taylor instability with possible implications for type Ia supernovae},
  author={Cabot, William H and Cook, Andrew W},
  journal={Nature Physics},
  volume={2},
  number={8},
  pages={562--568},
  year={2006},
  publisher={Nature Publishing Group UK London}
}

@article{zingale2005supernovae,
  title={Three-dimensional numerical simulations of Rayleigh-Taylor unstable flames in type Ia supernovae},
  author={Zingale, Michael and Woosley, SE and Rendleman, CA and Day, MS and Bell, JB},
  journal={The Astrophysical Journal},
  volume={632},
  number={2},
  pages={1021},
  year={2005},
  publisher={IOP Publishing}
}

@article{Clark2016NIF,
  author  = {Clark, D. S. and Weber, C. R. and Milovich, J. L. and Salmonson, J. D. and Kritcher, A. L. and Haan, S. W. and Hammel, B. A. and Hinkel, D. E. and Hurricane, O. A. and Jones, O. S. and Marinak, M. M. and Patel, P. K. and Robey, H. F. and Sepke, S. M. and Edwards, M. J.},
  title   = {Three-dimensional simulations of low foot and high foot implosion experiments on the {National Ignition Facility}},
  journal = {Physics of Plasmas},
  volume  = {23},
  number  = {5},
  pages   = {056302},
  year    = {2016},
  doi     = {10.1063/1.4943527}
}

@article{Zhou2019Mixing,
  author  = {Zhou, Ye and Clark, Timothy T. and Clark, Daniel S. and Glendinning, S. Gail and Skinner, M. Aaron and Huntington, Channing M. and Hurricane, Omar A. and Dimits, Andris M. and Remington, Bruce A.},
  title   = {Turbulent mixing and transition criteria of flows induced by hydrodynamic instabilities},
  journal = {Physics of Plasmas},
  volume  = {26},
  number  = {8},
  pages   = {080901},
  year    = {2019},
  doi     = {10.1063/1.5088745}
}

@article{Srinivasan2014Mix,
  author  = {Srinivasan, Bhuvana and Tang, Xian-Zhu},
  title   = {Role of hydrodynamic instability growth in hot-spot mass gain and fusion performance of inertial confinement fusion implosions},
  journal = {Physics of Plasmas},
  volume  = {21},
  number  = {10},
  pages   = {102702},
  year    = {2014},
  doi     = {10.1063/1.4897938}
}

@article{Brunton2016,
  title={Discovering governing equations from data by sparse identification of nonlinear dynamical systems},
  author={Brunton, Steven L and Proctor, Joshua L and Kutz, J Nathan},
  journal={Proceedings of the national academy of sciences},
  volume={113},
  number={15},
  pages={3932--3937},
  year={2016},
  publisher={National Academy of Sciences}
}

@article{Layzer1955,
  title={On the instability of superposed fluids in a gravitational field.},
  author={Layzer, David},
  journal={Astrophysical Journal, vol. 122, p. 1},
  volume={122},
  pages={1},
  year={1955}
}

@article{Goncharov2002,
  title={Analytical model of nonlinear, single-mode, classical Rayleigh-Taylor instability at arbitrary Atwood numbers},
  author={Goncharov, \_V N},
  journal={Physical review letters},
  volume={88},
  number={13},
  pages={134502},
  year={2002},
  publisher={APS}
}

@article{Kull1986,
  author  = {Kull, Hans-Joachim},
  title   = {Nonlinear Free-Surface Rayleigh--Taylor Instability},
  journal = {Physical Review A},
  volume  = {33},
  number  = {3},
  pages   = {1957--1967},
  year    = {1986},
  doi     = {10.1103/PhysRevA.33.1957}
}

@article{Zufiria1988,
  author  = {Zufiria, Juan A.},
  title   = {Bubble Competition in Rayleigh--Taylor Instability},
  journal = {Physics of Fluids},
  volume  = {31},
  number  = {3},
  pages   = {440--446},
  year    = {1988},
  doi     = {10.1063/1.866825}
}

@article{Mikaelian2003,
  author  = {Mikaelian, Karnig O.},
  title   = {Explicit Expressions for the Evolution of Single-Mode
             Rayleigh--Taylor and Richtmyer--Meshkov Instabilities
             at Arbitrary Atwood Numbers},
  journal = {Physical Review E},
  volume  = {67},
  number  = {2},
  pages   = {026319},
  year    = {2003},
  doi     = {10.1103/PhysRevE.67.026319}
}

@article{Liu2023,
  author  = {Liu, Changwen and Wu-Wang, Hongzhi and Zhang, Yousheng and Xiao, Zuoli},
  title   = {A Decoupled Mechanism of Interface Growth in Single-Mode Hydrodynamic Instabilities},
  journal = {Journal of Fluid Mechanics},
  year    = {2023},
  volume  = {964},
  pages   = {A37},
  doi     = {10.1017/jfm.2023.393}
}

@article{Liu2025,
  author  = {Liu, Changwen and Wu-Wang, Hongzhi and Zhang, Yousheng and Xiao, Zuoli},
  title   = {Prediction of long-time single-mode {Rayleigh--Taylor} bubble evolution: a rotational flow model},
  journal = {J. Fluid Mech.},
  year    = {2025},
  volume  = {1023},
  pages   = {A23},
  doi     = {10.1017/jfm.2025.10769}
}

@article{Piriz2024,
  author  = {Piriz, A. R. and L{\'o}pez Cela, J. J. and Piriz, S. A. and Tahir, N. A.},
  title   = {Nonlinear, single-mode, two-dimensional {Rayleigh--Taylor} instability in ideal media},
  journal = {Physical Review E},
  volume  = {110},
  number  = {2},
  pages   = {025101},
  year    = {2024},
  doi     = {10.1103/PhysRevE.110.025101}
}

@article{Piriz2026,
  author  = {Piriz, A. R. and Piriz, S. A. and Tahir, N. A. and Cobos, F.},
  title   = {Bubble and spike reacceleration in nonlinear, single mode, two-dimensional {Rayleigh--Taylor} instability},
  journal = {APS Open Sci.},
  year    = {2026},
  volume  = {1},
  pages   = {000149},
  doi     = {10.1103/px86-rwx6}
}

@article{Oron2001,
  title={Dimensionality dependence of the Rayleigh--Taylor and Richtmyer--Meshkov instability late-time scaling laws},
  author={Oron, Dan and Arazi, Lior and Kartoon, D and Rikanati, A and Alon, Uri and Shvarts, D},
  journal={Physics of Plasmas},
  volume={8},
  number={6},
  pages={2883--2889},
  year={2001},
  publisher={American Institute of Physics}
}

@article{Dimonte2000,
  title={Spanwise homogeneous buoyancy-drag model for Rayleigh--Taylor mixing and experimental evaluation},
  author={Dimonte, Guy},
  journal={Physics of plasmas},
  volume={7},
  number={6},
  pages={2255--2269},
  year={2000},
  publisher={American Institute of Physics}
}

@article{Rudy2017PDEFind,
  author  = {Rudy, Samuel H. and Brunton, Steven L. and Proctor, Joshua L. and Kutz, J. Nathan},
  title   = {Data-driven discovery of partial differential equations},
  journal = {Science Advances},
  volume  = {3},
  number  = {4},
  pages   = {e1602614},
  year    = {2017},
  doi     = {10.1126/sciadv.1602614}
}

@article{Champion2019Coordinates,
  author  = {Champion, Kathleen and Lusch, Bethany and Kutz, J. Nathan and Brunton, Steven L.},
  title   = {Data-driven discovery of coordinates and governing equations},
  journal = {Proceedings of the National Academy of Sciences},
  volume  = {116},
  number  = {45},
  pages   = {22445--22451},
  year    = {2019},
  doi     = {10.1073/pnas.1906995116}
}

@article{Brunton2020MLFluid,
  author  = {Brunton, Steven L. and Noack, Bernd R. and Koumoutsakos, Petros},
  title   = {Machine Learning for Fluid Mechanics},
  journal = {Annual Review of Fluid Mechanics},
  volume  = {52},
  pages   = {477--508},
  year    = {2020},
  doi     = {10.1146/annurev-fluid-010719-060214}
}

@article{Schaeffer2017,
  author  = {Schaeffer, Hayden and McCalla, Scott G.},
  title   = {Sparse model selection via integral terms},
  journal = {Physical Review E},
  volume  = {96},
  number  = {2},
  pages   = {023302},
  year    = {2017},
  doi     = {10.1103/PhysRevE.96.023302}
}

@article{Messenger2021,
  author  = {Messenger, Daniel A. and Bortz, David M.},
  title   = {Weak {SINDy}: {Galerkin}-based data-driven model selection},
  journal = {Multiscale Model. Simul.},
  volume  = {19},
  number  = {3},
  pages   = {1474--1497},
  year    = {2021},
  doi     = {10.1137/20M1343166}
}

\end{document}